\documentclass[twocolumn,amsmath,amssymb]{revtex4-1}

\usepackage[dvips]{graphicx}
\usepackage{dcolumn}
\usepackage[utf8]{inputenc}
\usepackage[T1]{fontenc}
\usepackage{mathptmx}
\usepackage{etoolbox}
\usepackage{siunitx}
\usepackage{color}
\usepackage{upgreek}
\usepackage{xspace}

\newenvironment{fequation*}{\empheq[box=\fbox]{equation*}}{\endempheq}
\newenvironment{falign*}{\empheq[box=\fbox]{align*}}{\endempheq}
\newenvironment{fmultline*}{\empheq[box=\fbox]{multline*}}{\endempheq}
\renewcommand{\bm}[1]{\boldsymbol{\mathbf{#1}}} \newcommand{\ud}{\mathrm{d}}

\renewcommand{\Re}{\operatorname{Re}}

\newcommand{\ie}{i.e.\@\xspace}

\usepackage[normalem]{ulem}

\begin{document}

\title{Recovering particle velocity and size distributions in ejecta with Photon Doppler Velocimetry}

\author{J.A.~Don Jayamanne}
\affiliation{CEA DIF, Bruyères-le-Châtel, 91297 Arpajon Cedex, France}
\affiliation{Institut Langevin, ESPCI Paris, PSL University, CNRS, 75005 Paris, France}
\author{R.~Outerovitch}
\affiliation{CEA DIF, Bruyères-le-Châtel, 91297 Arpajon Cedex, France}

\author{F.~Ballanger}
\author{J.~Bénier}
\author{E.~Blanco}
\affiliation{CEA DIF, Bruyères-le-Châtel, 91297 Arpajon Cedex, France}
\author{C.~Chauvin}
\author{P.~Hereil}
\affiliation{CEA, DAM, GRAMAT, BP 80200, F-46500 Gramat, France}
\author{J.~Tailleur}
\affiliation{CEA DIF, Bruyères-le-Châtel, 91297 Arpajon Cedex, France}

\author{O.~Durand}
\affiliation{CEA DIF, Bruyères-le-Châtel, 91297 Arpajon Cedex, France}
\affiliation{Paris-Saclay University, CEA, Laboratoire Matière en Conditions Extrêmes, 91180 Bruyères-le-Châtel, France}
\author{R.~Pierrat}
\affiliation{Institut Langevin, ESPCI Paris, PSL University, CNRS, 75005 Paris, France}
\author{R.~Carminati}
\email{remi.carminati@espci.psl.eu}
\affiliation{Institut Langevin, ESPCI Paris, PSL University, CNRS, 75005 Paris, France}
\affiliation{Institut d'Optique Graduate School, Paris-Saclay University, 91127 Palaiseau, France}
\author{A.~Hervouët}
\author{P.~Gandeboeuf}
\affiliation{CEA DIF, Bruyères-le-Châtel, 91297 Arpajon Cedex, France}
\author{J.-R.~Burie}
\email{jean-rene.burie@cea.fr}
\affiliation{CEA DIF, Bruyères-le-Châtel, 91297 Arpajon Cedex, France}

\date{\today}

\begin{abstract}
   When a solid metal is struck, its free surface can eject fast and fine particles. Despite the many
   diagnostics that have been implemented to measure the mass, size, velocity or temperature of ejecta, these
   efforts provide only a partial picture of this phenomenon. Ejecta characterization, especially in
   constrained geometries, is an inherently ill-posed problem. In this context, Photon Doppler Velocimetry
   (PDV) has been a valuable diagnostic, measuring reliably particles and free surface velocities in the
   single scattering regime. Here we present ejecta experiments in gas and how, in this context, PDV allows
   one to retrieve additional information on the ejecta, i.e. information on the particles’ size. We explain
   what governs ejecta transport in gas and how it can be simulated. To account for the multiple scattering of
   light in these ejecta, we use the Radiative Transfer Equation (RTE) that quantitatively describes PDV
   spectrograms, and their dependence on the velocity but also on the size distribution of the ejecta. We
   remind how spectrograms can be simulated by solving numerically this RTE and we show how to do so on
   hydrodynamic ejecta simulation results. Finally, we use this complex machinery in different ejecta
   transport scenarios to simulate the corresponding spectrograms. Comparing these to experimental results, we
   iteratively constrain the ejecta description at an unprecedented level. This work demonstrates our ability
   to recover particle size information from what is initially a velocity diagnostic, but more importantly it
   shows how, using existing simulation of ejecta, we capture through simulation the complexity of
   experimental spectrograms.
\end{abstract}

\maketitle

\section{Introduction}\label{introduction}

Probing matter's behavior under the extreme conditions of shock compression experiments allows one to better
understand its properties at rest. Ejecta formation, the process through which a shocked material ejects a
cloud of fast and fine particle, has been extensively studied lately~\cite{buttler_foreword_2017}. It has been
shown that ejecta is a limiting case of Richtmyer-Meshkov
instabilities~\cite{richtmyer_taylor_1960,meshkov_instability_1972} which occurs when the initial shockwave
interacts with the irregularities at the free surface of the
material~\cite{asay_ejection_1976,andriot_ejection_1982}. It causes matter to partially melt, creating
numerous expanding micro-jets. These micro-jets eventually fragment giving birth to the actual ejecta [see
Fig.~\ref{setup_ejection}(b)]. One of the purpose of ejecta study is to determine the size-velocity
distribution of this particle cloud.

Advances on the ejecta source model
theory~\cite{buttler_unstable_2012,dimonte_ejecta_2013,georgievskaya_model_2017} in shock compression
experiments and the corresponding
simulation~\cite{fung_ejecta_2013,durand_large-scale_2012,durand_power_2013,durand_mass-velocity_2015} have
permitted a better description of the distribution created by a given sample in response to a given
solicitation. These simulations study the particles from their creation at the early moment of the experiment
to their transport throughout the propagation medium. This simulation effort on the ejecta side was supported
by the experimental development of numerous and diverse optical diagnostics which refined ejecta description
to further constrained ejecta simulations. Especially, Mie
scattering~\cite{monfared_ejected_2015,schauer_ejected_2017} and holography
diagnostics~\cite{sorenson_ejecta_2014,guildenbecher_ultraviolet_2023} have given valuable insights on the
particle size-velocity distribution. The main limitations of these diagnostics remain their difficulty of
implementation and the fact that they only allow the study of elementary examples of ejecta formation, namely
ejection with a few micro-jets. For now, one has no other choice than assuming that the ejection process in
more complex experiments gives the same particle size-velocity distribution with no mean of verifying this
claim.

Photon Doppler Velocimetry~(PDV) is another optical diagnostic which was initially developed to monitor
particle velocity distributions~\cite{STRAND-2006,mercier_photonic_2006}. With a single scattering hypothesis,
the PDV response of an ejecta, its time-velocity spectrogram, can be seen as the velocity distribution of the
ejecta at a given time. Recently, we have shown~\cite{don_jayamanne_characterization_2024} that this
spectrogram is in fact the solution of a broader light transport model which is sensitive to the particle size
distribution and its statistical inhomogeneities throughout the medium. Compared to Mie scattering and
holography diagnostics that are based on off-axis and transmission measurements, PDV is on-axis and in
reflection. A unique collimated probe is used for illumination and detection. This makes it compact, reliable
and minimum invasive. These perks have made it one of the key diagnostic implemented in almost all experiments
and especially the most constrained ones. Showing that it is possible to recover additional size information
from a PDV spectrogram would highly impact ejecta analysis. This would enable the evaluation of particle sizes
in the most complex configurations and allow one to verify that the ejecta formation process is in fact
similar to that of elementary experiments.

The purpose of this work is to showcase an experiment where, using only PDV spectrograms, we constrain the
size distribution of an ejecta. To proceed, we first introduce ejecta experiments in gas, their interest for
particle size study and the means available to simulate particle transport in such media. Then, we remind the
working principle of PDV measurements, why they are sensitive to the ejecta size distribution and how we can
compute simulated spectrograms out off of simulated ejecta transport. Matching the simulated spectrograms to
the experimental ones acquired for three different gas conditions, we are able to better constrain the initial
size distribution of the ejecta and test its robustness. Finally, we discuss how the present work reflects on
existing literature and ejecta understanding.

The paper is organized as follows. Section~\ref{ejecta} is dedicated to the presentation of ejecta experiments
in gas. We introduce the micro-jetting mechanism and the resulting size-velocity distribution. We explain the
ways ejecta is assumed to interact with gas and how it depends on particle size. We then introduce the
\emph{Ph\'enix} code that handles ejecta transport in gas and allows one to compute the expected ejecta
description at each step of the corresponding experiment. In Sec.~\ref{pdv}, we recall the PDV instrumentation
ejecta experiments receive and how, with a single scattering hypothesis, the resulting PDV spectrogram
accurately estimates the velocity distribution of the ejecta. For the ejecta considered here, multiple
scattering cannot be ignored. In this regime, we rewrite this spectrogram as a function of the specific
intensity, a commonly used quantity in statistical optics. This quantity is the solution of a Radiative
Transfert Equation (RTE) that has been modified to account for both Doppler shifts due to the particles
movement and the statistical inhomogeneities of the ejecta. The PDV spectrogram can then be computed given the
ejecta's size-velocity distribution in time and we do so directly on ejecta transport results obtained with
the \emph{Ph\'enix} hydrodynamic simulation code. Our comprehensive study of particle size distribution based
on simulated PDV spectrograms in different gas is reported in Sec.~\ref{results}. We start in the simplest
case of vacuum and then gradually increase the complexity with helium and air. Finally, Sec.~\ref{discussion}
discusses the implications of this work, how it compares to existing literature and what it argues in favor of
for future ejecta studies.

\section{Ejecta experiments and simulation in gas}\label{ejecta}

\subsection{Ejecta creation}\label{ejecta_creation}

\begin{figure}[!htb]
   \centering
   \includegraphics[width=0.80\linewidth]{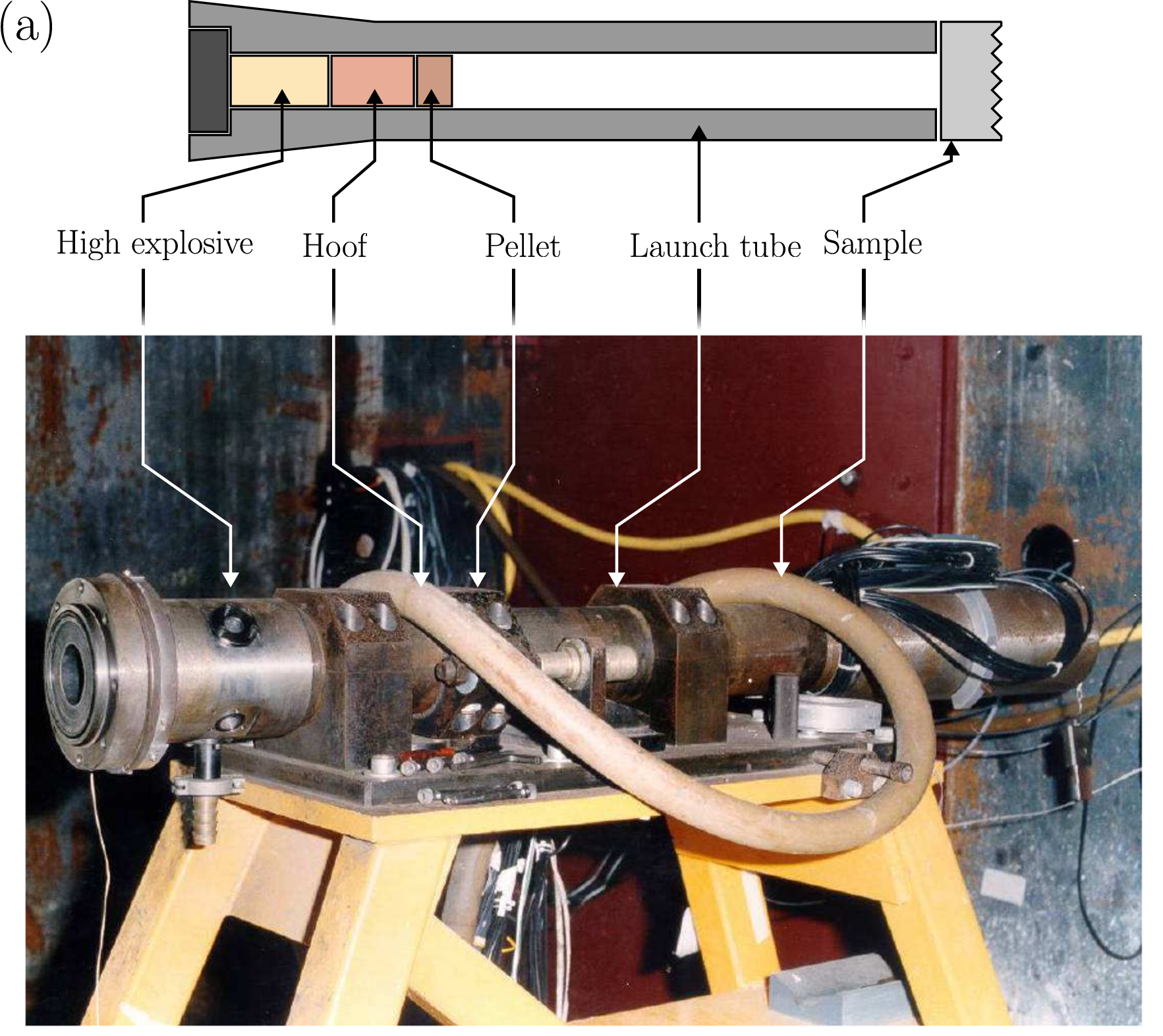}
   \includegraphics[width=0.80\linewidth]{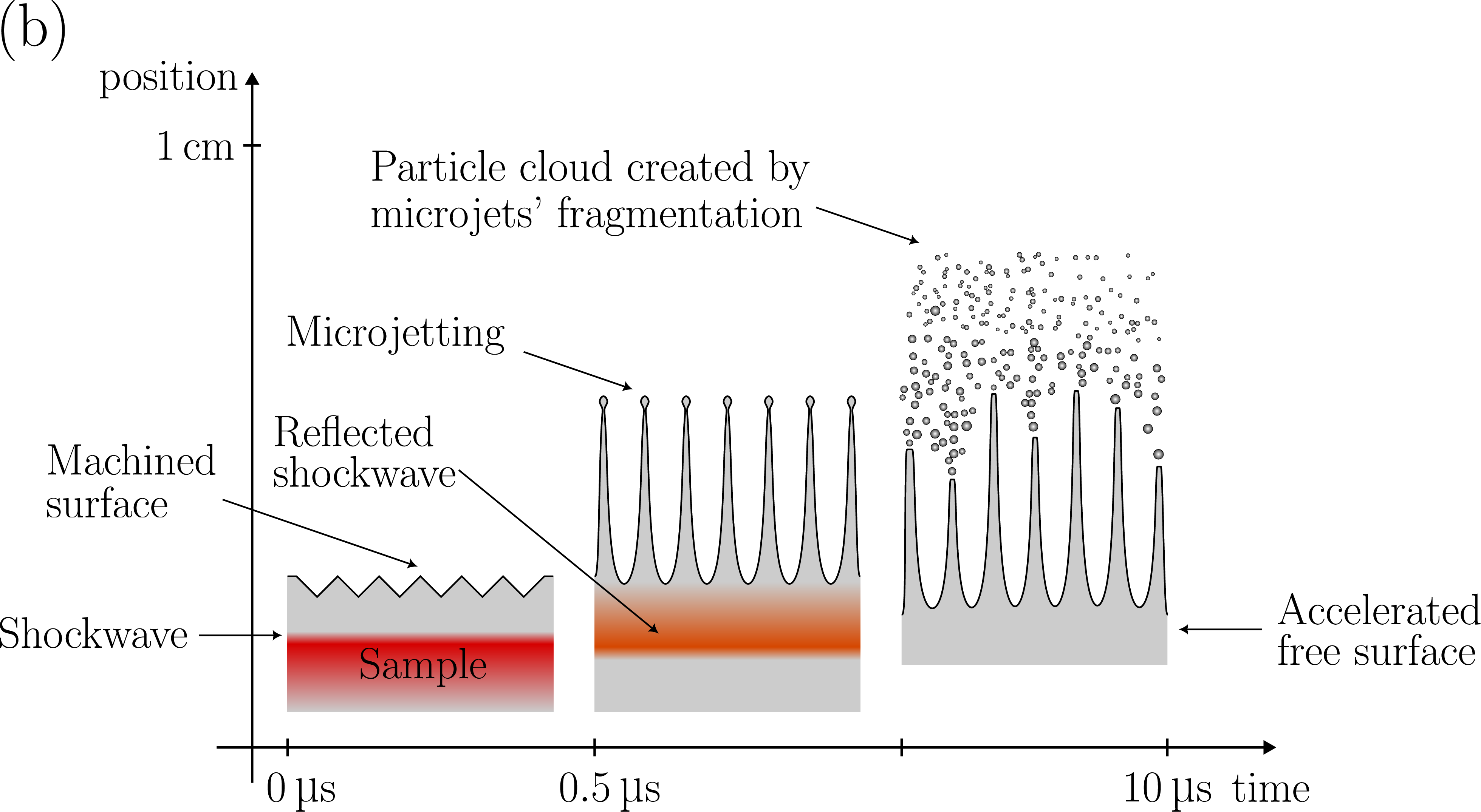}
   \caption{(a)~Typical explosive setup of a planar shock experiment. (b)~Illustration of the micro-jet
   mechanism in a typical shock ejecta experiment. Upon reaching the machined free surface, the shock wave
   first comes into contact with the inwardly directed grooves. Under right angle conditions, the shock wave
   is reflected and the inward grooves become outward micro-jets. Due to the velocity gap between the
   jet-heads and the free surface, the micro-jets are stretched until surface tension is no longer sufficient
   to hold matter together and fragmentation begins. This results in the creation of an ejecta.}
   \label{setup_ejection}
\end{figure}

Typical ejecta experiments, as the ones studied in Sec.~\ref{results}, are the planar shock experiments. In a
tube, a sample of the material of interest, here a grooved surface of tin~(Sn), is shocked by a High-Explosive
(HE) driven pellet [see Fig.\ref{setup_ejection}(a)]. The shockwave's interaction with the surface
irregularities creates liquid micro-jets of matter. These micro-jets eventually undergo fragmentation giving
birth to the actual ejecta. In our experiment, the barrel's inner diameter is
$\Phi_{\text{barrel}}=\SI{98}{\milli m}$, the samples are tin disks with $\SI{60}{\micro m} \times
\SI{8}{\micro m}$ surface groves. With a copper flyer hitting the tin samples at $\SI{1650}{m/s}$, we reach a
shock pressure of $P_{\text{shock}}=\SI{29.5}{GPa}$, which ensures liquid phase transition in expansion.

To model this ejecta in gas, we need to make a few assumptions. The first one is that the ejecta is made out
of spherical particles with radius $a$. Right after impact, it can then be described by its initial
size-velocity distribution $g(a,\bm{v})$ normalized such that
\begin{equation}
   \int g(a,\bm{v}) \ud a \ud \bm{v}=1\,.
\end{equation} 
While recent holography imaging results~\cite{sorenson_ejecta_2014,guildenbecher_ultraviolet_2023} and
molecular dynamics simulations~\cite{durand_large-scale_2012,durand_power_2013,durand_mass-velocity_2015}
suggest otherwise, this is discussed in Sec.~\ref{discussion}, our second assumption is that the initial size
and velocity distributions are independent. This reads
\begin{equation}
   g(a,\bm{v})=h(a)j(\bm{v})\, ,
\end{equation}
with $h(a)$ the size distribution, $j(\bm{v})$ the velocity distribution and
\begin{align}
   &\int h(a) \ud a =1 \, ,
\\
   &\int j(\bm{v}) \ud \bm{v} =1\,.
\end{align}
Our last assumption is that the initial size and velocity distributions do not depend on the presence nor the
nature of the gas in the chamber. Once again Mie-scattering measurements have shown the limits of such an
hypothesis~\cite{buttler_understanding_2021}, but as a first approximation it will prove useful.

In Sec.~\ref{results}, different initial size distributions will be tested but the velocity distribution will
be imposed from here on. Instead of defining it directly, we prefer using the integrated mass-velocity
distribution $M(v)$ where $v=|\bm{v}|$. Assuming all velocities to be along the ejection direction $\bm{u}_z$,
the distribution then reads
\begin{equation}
   j(\bm{v})=-\frac{1}{M_s}\frac{\ud}{\ud v}M(v)\delta\left(\frac{1}{v}\bm{v}-\bm{u}_z\right)\, ,
\end{equation}
where $M_s$ is the surface mass. Independent Asay foil measurements~\cite{asay_ejection_1976} in vacuum for
these experiments give
\begin{equation}
M(v)=
\begin{cases}
   M_s\exp\left[-\beta\left(\frac{v}{v_s}-1\right)\right],&\text{if}\,v \in [v_\text{min},v_\text{max}]\\
   0,&\text{otherwise}
\end{cases}
\end{equation}
with the surface mass $M_s=\SI{12}{\milli g / \centi m^2}$, $\beta=\SI{11.7}{}$,
$v_s=v_\text{min}=\SI{2060}{m/s}$ and $v_\text{max}=\SI{3350}{m/s}$.

Now that the initial properties of the ejecta have been described, we need to model its interaction with gas.

\subsection{Ejecta transport in gas}\label{ejecta_transport}
In case of a vacuum-tight chamber, the objects are in ballistic transport and the ejecta size-velocity
distribution remains the same during propagation. Now if the chamber is gas filled, the transport properties
of the ejecta are altered. While we assume the same initial size-velocity distribution as in vacuum, knowing
that gas interaction depends both on particle size and velocity, the size-velocity distribution will evolve
with time. This discrimination is what makes ejecta experiments in gas extremely insightful. The same set of
initial conditions in different gases must allow us to retrieve radically different particles transport
scenarios. In other words, the robustness of a unique initial description can be tested against different gas
transport conditions to see whether or not it allows one to match all experimental results. While different
from a direct measurement, this technique must be seen as a powerful new way of evaluating particle
size-velocity distribution with existing diagnostics and in otherwise inaccessible configurations.

In the presence of gas in front of the ejecta, the particles interact with it mainly through the drag force,
which tends to drive the particle toward the gas velocity. The drag force depends on the particle radius, the
gas density and the drag coefficient (based on the particle's Reynolds and Mach numbers). We take into account
the two-way coupling of the particle and the gas. The drag force is calculated using the KIVA-II
formulation~\cite{Amsden1989} given by,
\begin{equation}
  \label{eq:drag}
  \mathbf{F}_p = -\frac{1}{2} \pi {a_p}^2 \rho_g  \left(\bm{v}_p - \bm{v}_g\right) \left|\bm{v}_p - \bm{v}_g\right| C_d,
\end{equation}
with $a_p$ the particle radius, $\rho_g$ the gas density, $\bm{v}_g$ and $\bm{v}_g$ the gas and particle
velocity and $C_d$ the drag coefficient.

A hydrodynamic break-up model is also introduced. Particles can break-up according to their Weber number
$\text{We}_p$ and give birth to smaller-sized particles. The Weber number is expressed as the ratio between
hydrodynamic forces and the surface tension of the particle,
\begin{equation}
  \label{eq:weber}
  \text{We}_p = \frac{ 2 a_p \rho_g \left|\bm{v}_p - \bm{v}_g\right|^2}{\sigma_p},
\end{equation}
with $\sigma_p$ the particle surface tension.

Given an initial ejecta distribution and these two interactions, we need a simulation handling the transport
of particles accordingly in time.

\subsection{\emph{Ph\'enix} code for particle transport simulation}\label{ejecta_simulation}
Hydrodynamic simulations have become the standard to compute matter's behavior in shock compression
experiments~\cite{fung_ejecta_2013}. In this work, the simulations are run with the \emph{Ph\'enix} code,
developed at CEA which uses a multiphase particulate transport method to model two-way coupling of momentum
and energy. This is based on the approach proposed by Amsden~\emph{et al.} and implemented in the KIVA-II
code~\cite{Amsden1989}, which has been improved from the original paper.

To perform the simulations, we have to initialize the particle cloud according to the experimental parameters
to fit the other diagnostics implemented in the experimental set-up. We define the ejected mass velocity curve
$M(v)$ in agreement with the Asay foil~\cite{asay_ejection_1976} measurement under vacuum (still assuming the
gas does not change the total ejected mass). For the initial size distribution, we rely on previous
experiments and deal with power laws or lognormal distributions. Studying its influence on the spectrograms is
one of the purpose of this article as described in Sec.~\ref{results}.

For each experiment, the \emph{Ph\'enix} code gives the corresponding cloud description at different time
steps. For each of these times, the description corresponds to a list of so-called \emph{numerical particles}.
Instead of actually simulating the transport of each individual particles, we only consider a smaller set of
numerical particles. Each of them has a size, a numerical weight, \ie the number of physical particles it
represents, a position and a velocity. The number of numerical particles must be high enough to encompass the
full dynamics of the ejecta while being low enough to ensure a reasonable compute time. On the one hand, the
drag force affects the velocity and therefore the position of the numerical particles. On the other hand, the
break-up process, which takes place only if the Weber number $We_p$ is superior to the critical Weber number
$We_\text{crit} = 15$, decrease the particle size. In order to respect mass conservation, this decrease in
size is coupled to an increase of the numerical weight assigned to the particle. As an order of magnitude,
each ejecta simulation corresponding to Sec.~\ref{results} have around $2500$ numerical particles, taken at
$180$ temporal steps each separated by $\delta t=\SI{0.16}{\micro s}$. The corresponding compute time is
$\SI{2}{h}$ on 1 AMD EPYC 7763 64-core CPUs clocked at $\SI{2.45}{\giga Hz}$.

In summary, the three main parameters that control the evolution of the particle cloud during the simulation
are the initial size distribution, drag force and break-up model. In the study reported in Sec.~\ref{results},
these parameters are the ones we expect to fine tune to make the ejecta robust to transport in different
gases. The efficiency of such a procedure depends on the sensitivity of the chosen diagnostic to changes in
ejecta transport properties. The diagnostic chosen here, PDV, is presented in Sec.~\ref{pdv}.

\section{Photon Doppler Velocimetry in ejecta}\label{pdv}

While PDV in ejecta has been shown to be deep in the multiple
scattering~\cite{don_jayamanne_characterization_2024} regime (which will be the case for all the ejecta
presented here), it is interesting to consider first the single scattering regime for which PDV was initially
developed. This is the purpose of Sec.~\ref{pdv_single}.

\subsection{Photon Doppler Velocimetry in the single scattering regime}\label{pdv_single}
\begin{figure}[!htb]
   \centering
   \includegraphics[width=0.80\linewidth]{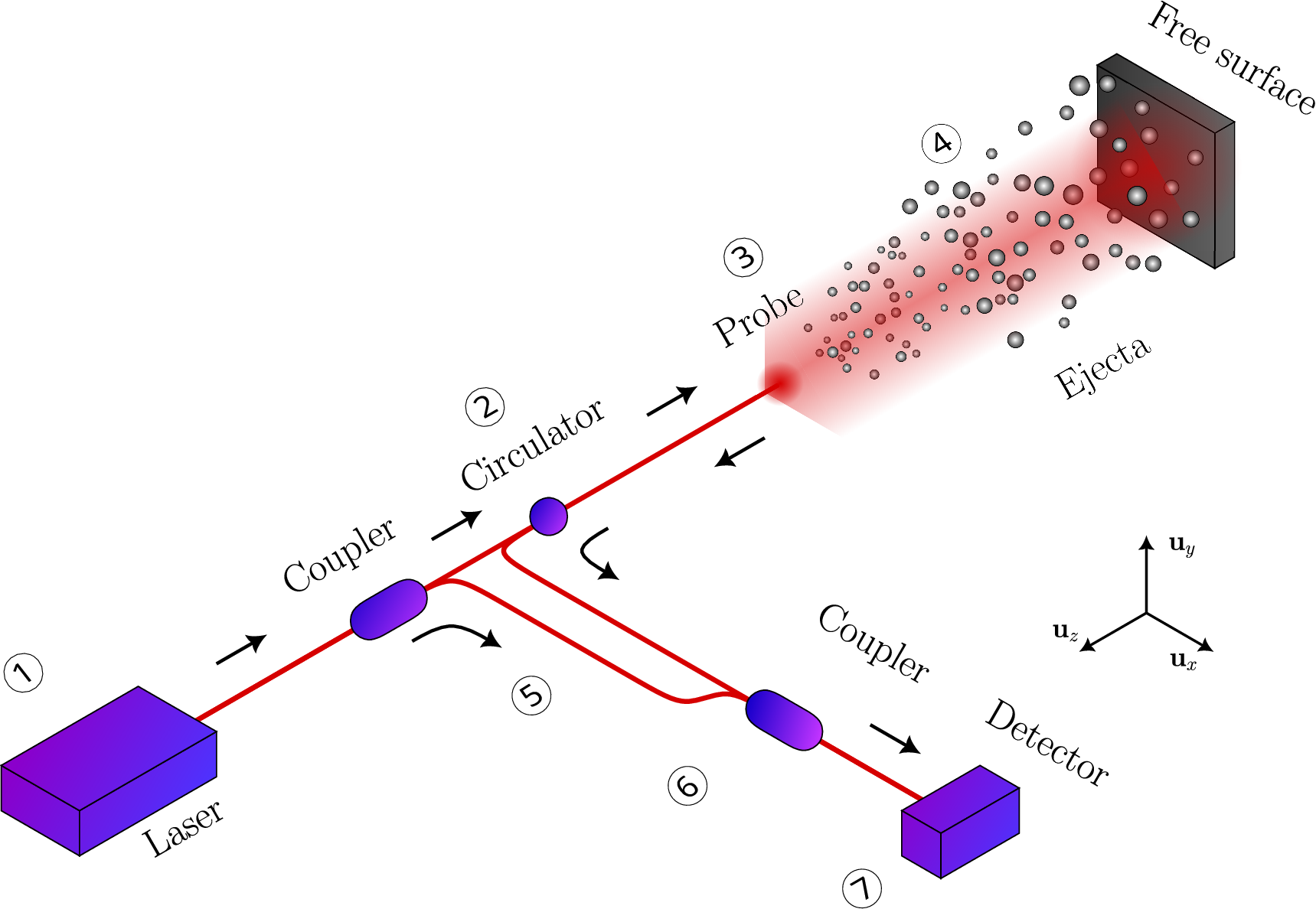}
   \caption{Schematic representation of a typical shock-loaded experiment with a PDV setup. The probe
   illuminates the ejecta and the free surface with a highly collimated laser beam (numerical aperture of
   $\SI{4.2}{\milli rad}$ and pupil size $\phi_p = \SI{1.3}{\micro m}$). The backscattered field is collected
   by the probe acting as the measuring arm and interferes with the reference arm at the detector. The beating
   signal is registered with a high bandwidth oscilloscope before being analyzed.}
   \label{setup_pdv}
\end{figure}

Photon Doppler Velocimetry is an interferometric technique~\cite{STRAND-2006,mercier_photonic_2006} where a
collimated laser beam at frequency $\omega_0$ is shined toward a cloud of moving particles and a free surface.
As seen in Fig.~\ref{setup_pdv}, light then get scattered by this ejecta and slightly shifted in frequency
before part of it is captured in reflection. The collected field interferes at the detector with a reference
field at $\omega_0$, resulting in a beating signal $\mathcal{I}(t)$ at the detector. This signal can be
written
\begin{equation}\label{useful_detected_signal}
   \mathcal{I}(t)=\int2\Re\left[\bar{E}_s(\bm{r},t)\bar{E}_0^*(\bm{r},t)\right]\ud \bm{r}\, ,
\end{equation}
with $\bar{E}_s(\bm{r},t)$ the analytic signal associated to the scattered field, $\bar{E}_0(\bm{r},t)$ the
analytic signal associated to the reference field and where $\ud\bm{r}$ denotes integration over the detector
surface.

In post-treatment a Short-Term Fourier Transform (STFT) is applied defining the spectrogram $S(t,\omega)$ as
\begin{equation}\label{spectrogram}
   S(t,\omega)=\left|\int \mathcal{I}(\tau)w(\tau-t)\exp(i\omega\tau)\ud \tau\right|^2
\end{equation}
where $w(t)$ is a gate function of typical width $T_w$ such that $\int w(t)\ud t=T_w$.

With a single scattering hypothesis, the scattered field is the sum of the fields scattered by each particle.
We consider scalar fields, since our focus in this work will be in the multiple scattering regime, where the
field can be considered unpolarized, since depolarization is known to occur on scales of the order of the
scattering mean free path~\cite{VYNCK-2014}. For a number of particle $N(t)$, and assuming a detection in the
far field, the scattered field then reads
\begin{multline}\label{scattered_field_single}
   \bar{E}_s(\bm{r},t)=\frac{e^{ik_0r}}{r}\sum_{j=1}^{N(t)}\mathcal{A}_j(\bm{u},\bm{u}_0,t)
\\\times
   \exp\left\{-i\left[\omega_0
   +k_0(\bm{u}-\bm{u}_0)\cdot\bm{v}_j(t)\right]t\right\}\,,
\end{multline}
where $r=|\bm{r}|$, $k_0=\omega_0/c$ with $c$ the light velocity in vacuum, $\bm{u}_0$ is the unit vector
defining the direction of illumination, $\bm{u}=\bm{r}/r$ defines the direction of observation,
$\mathcal{A}_j(\bm{u},\bm{u}_0,t)$ is the amplitude of the field scattered by particle $j$ and $\bm{v}_j(t)$
its velocity. Using Eq.~(\ref{scattered_field_single}) and for an observation direction $\bm{u}=-\bm{u}_0$,
the spectrogram given in Eq.~(\ref{spectrogram}) becomes 
\begin{multline} \label{spectrogram_single}
   S(t,\omega)=\frac{\pi^2\left|\mathcal{A}_0\right|^2}{r^2}\int\sum_{j=1}^{N(t)}\left|\mathcal{A}_j(-\bm{u}_0,\bm{u}_0,t)\right|^2
\\\times
   \left\{\left|\delta\left[\omega+\frac{4\pi}{\lambda} v_j(t)\right]\right|^2+\left|
   \delta\left[\omega-\frac{4\pi}{\lambda} v_j(t)\right]\right|^2\right\}\ud\bm{r}\, ,
\end{multline}
where $\mathcal{A}_0$ is the amplitude of the reference field and $\delta$ the Dirac delta function. This
simple expression allows one to convert the frequency appearing in PDV spectrograms directly into a velocity
using $v=\omega/(4\pi)\lambda$. In this regime, as shown in Fig.~\ref{spectrogram_comparison_vacuum_power}(a),
the spectrogram gives an accurate estimation of the velocity distribution in the ejecta.

\subsection{Photon Doppler Velocimetry beyond single scattering}\label{pdv_multiple}

With $z$ denoting the depth along the ejection direction, the single scattering hypothesis holds as long as
the optical thickness $b\ll 1$. The optical thickness is defined as
\begin{equation}\label{b}
   b=\int\frac{1}{\ell_s(z)}\ud z
\end{equation}
where $\ell_s(z)$, the photon scattering mean-free path, will be introduced in greater details in this
section. We have shown that in ejecta experiments $b$ can far exceed unity, for example $b=42$ in the study by
Shi~\emph{et al.}~\cite{shi_reconstruction_2022} In these conditions, Eq.~(\ref{scattered_field_single}) does
not hold and the spectrogram expression must be enriched to account for multiple scattering. This was the
purpose of a previous work~\cite{don_jayamanne_characterization_2024} where we explain in great details and
with the relevant hypothesis how the PDV spectrogram expression given in Eq.~(\ref{spectrogram_single}) can be
extended to the multiple scattering regime. The purpose of this section is to recall the most important
results of this previous work that will be useful to perform spectrogram simulation in Sec.~\ref{results}.

The quantity of interest in the multiple scattering regime is the specific intensity
$I(\bm{r},\bm{u},t,\omega)$~\cite{BARABANENKOV-1969,RYTOV-1989,APRESYAN-1996,carminati_principles_2021}. This
radiometric quantity can be interpreted as a radiative flux at position $\bm{r}$, in direction $\bm{u}$, at
time $t$ and at frequency $\omega$. In this sense, we can show that the specific intensity satisfies the
Radiative Transfer Equation (RTE) which will be presented in detail below. The specific intensity can also be
related to the wave field via the Fourier transform of its correlation function. This definition allows one to
connect the specific intensity to the spectrogram by the relation
\begin{multline}\label{spectrogram_full}
   \delta(k-k_r)S(t,\omega)
   =T_w\left|\mathcal{A}_0\right|^2
\\\times   
   \int_G\left[I(\bm{r},\bm{u},t,\omega_0+\omega)+I(\bm{r},\bm{u},t,\omega_0-\omega)\right]  
   \bm{u}\cdot\bm{n}
   \ud\bm{u}\ud\bm{r} \, ,
\end{multline}
where $k_r=n_r\omega_0/c$, $n_r$ being the real part of $n_\text{eff}$ the effective refractive index of the
medium as defined in Ref.~\onlinecite{don_jayamanne_characterization_2024}, $G$ is the etendue of the detector (surface of detection and angular aperture), $\ud \bm{u}$
corresponds to integration over the solid angle, and $\bm{n}$ is the unit vector normal to the detector
surface.

The RTE governing the evolution of the specific intensity takes the
form~\cite{don_jayamanne_characterization_2024}
\begin{multline}\label{rte}
   \left[\frac{1}{v_E(\bm{r},t,\omega)}\frac{\partial}{\partial t}+\bm{u}\cdot\bm{\nabla}_{\bm{r}}+\frac{1}{\ell_e(\bm{r},
   t,\omega)}\right]
   I(\bm{r},\bm{u},t,\omega)\\
   =\frac{1}{\ell_s(\bm{r},t,\omega)}\int p(\bm{r},\bm{u},\bm{u}',t,\omega,\omega')
   I(\bm{r},\bm{u}',t,\omega')\ud\bm{u}'\frac{\ud \omega'}{2\pi}\,,
\end{multline}
with $v_E$ the energy velocity, $\ell_e$ the extinction mean-free path and $p$ the phase function.
Equation~(\ref{rte}) is a generalized form of RTE that takes into account the inhomogeneities of the particle
cloud (under a quasi-homogeneous approximation~\cite{MANDEL-1995,hoskins_radiative_2018}). This equation that
naturally accounts for multiple scattering can be understood as an energy balance. The two derivatives of the
specific intensity in the left-hand side of Eq.~(\ref{rte}) corresponds to the spatio-temporal evolution of
this quantity. This evolution is governed by both losses and gains. Losses are caused by absorption and
scattering as described by the extinction mean-free path $\ell_e$. It is worth pointing out that these losses
happen at the same frequency $\omega$. The gains, also caused by scattering, are handled by the phase function
in the right-hand side of Eq.~(\ref{rte}). The scattering process being inelastic, it allows a conversion from
a frequency $\omega'$ to $\omega$. The extinction mean-free path $\ell_e$ is defined as
\begin{equation}\label{l_e}
   \frac{1}{\ell_e(\bm{r},t,\omega)}
      =\int\rho(\bm{r},t) \sigma_{e}(a,\omega)h(\bm{r},t,a)\ud a,
\end{equation}
where $\sigma_{e}(a,\omega)$ is the extinction cross-section of a particle with radius $a$ at frequency
$\omega$ and $h(\bm{r},t,a)$ is the size distribution at position $\bm{r}$ and time $t$. In our case
$h(\bm{r},t_0,a)=h(a)$ as introduced in Sec.~\ref{ejecta}, with $t_0$ the ejecta creation time. The scattering
mean-free path $\ell_s$ and the phase function $p$ are defined as
\begin{multline}\label{l_s_p}
    \frac{1}{\ell_s(\bm{r},t,\omega)} p(\bm{r},\bm{u},\bm{u}',t,\omega,\omega')
    =\int\rho(\bm{r},t) \frac{\ud\sigma_{s}(a,\bm{u}\cdot\bm{u}',\omega)}{\ud\bm{u}}
\\\times
    2\pi\delta\left[\omega'-\omega-k_R(\bm{u}'-\bm{u})\cdot\bm{v}\right]
    g(\bm{r},t,a,\bm{v})\ud a\ud\bm{v},
\end{multline}
where $\sigma_{s}(a,\omega)$ is the scattering cross-section of a particle with radius $a$ and
$g(\bm{r},t,a,\bm{v})$ is the size-velocity distribution at position $\bm{r}$ and time $t$. Again, in our
case, $g(\bm{r},t_0,a,\bm{v})=h(a)j(v)$ as introduced in Sec.~\ref{ejecta}. With this definition, the phase
function is normalized as
\begin{equation}
   \int p(\bm{r},\bm{u},\bm{u}',t,\omega,\omega')\ud \bm{u}'\frac{\ud\omega'}{2\pi}=1\,,
\end{equation}
and, integrating Eq.~(\ref{l_s_p}) over $\bm{u}$, the scattering mean free-path reads
\begin{equation}\label{l_s}
   \frac{1}{\ell_s(\bm{r},t,\omega)}
      =\int \rho(\bm{r},t) \sigma_{s}(a,\omega)h(\bm{r},t,a)\ud a\,.
\end{equation}
We define the absorption mean-free path $\ell_a(\bm{r},t)$ as
\begin{equation}
   \frac{1}{\ell_a(\bm{r},t)}=\frac{1}{\ell_e(\bm{r},t)}-\frac{1}{\ell_s(\bm{r},t)}.
\end{equation}
Finally, since we have nonresonant scattering, the energy velocity $v_E$ is given by $v_E=c/n_\text{eff}$.

The quasi-homogeneous approximation used in this form of the RTE, which amounts here to the $\bm{r}$ and $t$
dependencies of the ejecta description, makes it perfectly suited here to evaluate the effects of
inhomogeneities caused by ejecta experiments in gas. Moreover, the dependance of the mean-free paths and phase
function on the size-velocity distribution of the ejecta $g(\bm{r},t, a, \bm{v})$ makes PDV a good candidate
for indirect size-velocity measurement we aim to perform. While the idea of ejecta experiments in gas to
evaluate particle size-velocity distribution is not new~\cite{remiot_etude_1991, buttler_understanding_2021},
the novelty of this work resides in connecting historic ejecta simulations schemes directly to a light
transport model for PDV. The intricacies of this link is the subject of Sec.~\ref{pdv_simulation}.

\subsection{Photon Doppler Velocimetry spectrogram simulation}\label{pdv_simulation}

The RTE given in Eq.~(\ref{rte}) can be rewritten in an integral form which is naturally suited for a
Monte~Carlo simulation scheme. Such a process can be seen as a random walk for energy quanta (behaving as
classical particles), where each step is sampled in statistical distributions for step length, scattering
direction, and frequency. While the expression of these statistical distributions and the details of the
random walk procedure are available in Ref.~\onlinecite{don_jayamanne_characterization_2024}, what is relevant
in the scope of this paper is that they only require the mean-free paths and phase function defined in
Eqs.~(\ref{l_e}) and~(\ref{l_s_p}) on a time scale small enough to seize de temporal evolution of the ejecta
and on a spatial scale small enough to capture its spatial statistical inhomogeneities. The challenge here is
to connect the ejecta description, \ie the numerical particles, computed by the \emph{Ph\'enix} code to the
aforementioned quantities of interest.

\begin{figure}[!htb]
   \centering
   \includegraphics[width=0.8\linewidth]{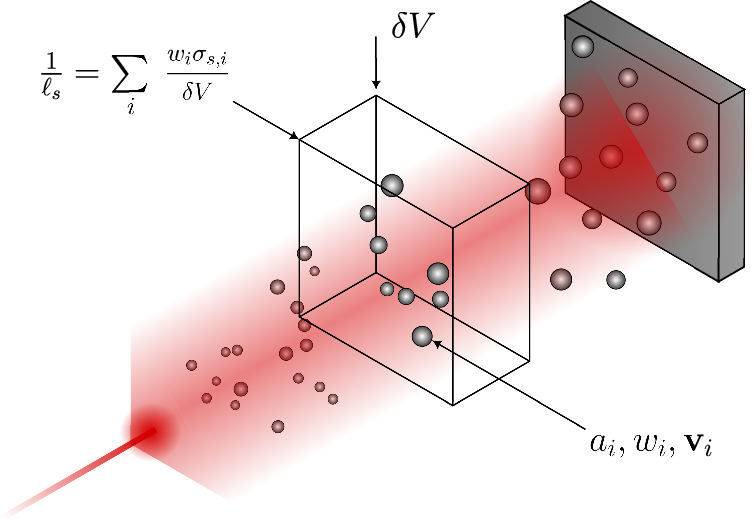}
   \caption{Illustration of how to use hydrodynamic simulation results as input data for PDV spectrogram
   simulation. The medium is sliced in layers of equal thickness and each of them intersects part of the
   ejecta. For each layer, the numerical particles concerned are then used to compute the local mean-free
   paths and phase functions.}
   \label{hydro_to_mc}
\end{figure}

Firstly, the value of $\delta t$, the time step introduced in Sec.~\ref{ejecta} chosen for the \emph{Ph\'enix}
code, is well below the typical value of $\SI{1}{\micro s}$ for the ejecta evolution time scale. Same goes for
the spatial scale where the typical number of numerical particles of $2500$ mentioned in Sec.\ref{ejecta}
provides a good sampling of the ejecta on position, velocities and size. Now, at a given step of the
hydrodynamic simulation, here is how the resulting ejecta description, \ie the numerical particles and their
attributes, allow to compute the corresponding mean-free paths and phase function. The numerical particles are
given in the launch tube geometry. This ejecta is then discretized spatially into several layers depending on
the space variations of the ejecta's statistical properties. In each layer the integration over size $a$ and
velocity $\bm{v}$ appearing in Eqs.~(\ref{l_e}) and~(\ref{l_s_p}) is replaced by a discrete sum over the
numerical particle. In these discrete sums, the product $\rho(\bm{r},t)h(\bm{r},t,a)$ and
$\rho(\bm{r},t)g(\bm{r},t,a,\bm{v})$ is replaced by the particle number density $w_i/\delta V$ where $w_i$
corresponds to the numerical weight of the $i^\text{th}$ numerical particle and $\delta V$ the volume of the
layer in the ejecta geometry. The regular and differential cross-sections are computed using the routine given
in Ref.~\onlinecite{bohren_absorption_1983}. An example of this entire procedure is depicted in
Fig.~\ref{hydro_to_mc} for computing the extinction mean-free path.
As an order of magnitude, the compute time for each simulated spectrogram presented in Sec.~\ref{results} is
around $\SI{1}{h}\,\SI{20}{min}$ on 80 AMD EPYC 7763 64-core CPUs clocked at $\SI{2.45}{\giga Hz}$ for
$\SI{5.12e+9}{}$ Monte~Carlo draws, at $180$ different times, with $2500$ numerical particles arranged in
$100$ effective layers.

After presenting the simulation tools permitting to describe the path from the solicitation on the sample to
the spectrogram, the aim of Sec.~\ref{results} is to do a comprehensive study to constrain ejecta description
using simulated PDV spectrograms.

\section{Ejecta behavior recovery based on PDV spectrograms analysis}\label{results}
 
In the remainder of this article, we propose to study a complex set of ejection experiments in gas. We
consider three experiments, differing only by the gas present in the chamber. The experimental setup is the
one presented in Sec.~\ref{ejecta_creation} and pictured in Fig.~\ref{setup_ejection}. Ejecta travels,
respectively, in $P_\text{vacuum}=\SI{e-5}{bar}$ vacuum, $P_\text{helium}=\SI{5}{bar}$ helium and
$P_\text{air}=\SI{1}{bar}$ air.

We start by simulating a spectrogram in vacuum to define a size-velocity distribution baseline. Next, in order
to see if this baseline holds in helium, we focus on the induced drag forces. Finally in air, we explore the
additional effect of hydrodynamic break-up.

\subsection{Ejecta's initial size distribution in vacuum}\label{vacuum}

For this first simulation in vacuum, we choose a standard power law distribution of the particle size, in the
form~\cite{schauer_ejected_2017}
\begin{equation}
   h(a)=
   \begin{cases}
      \frac{\alpha-1}{a_\text{min}^{-\alpha+1}-a_\text{max}^{-\alpha+1}} a^{-\alpha},&\text{if}\,a \in [a_\text{min},a_\text{max}]\\
      0,&\text{otherwise}
   \end{cases}
   \,,
\end{equation}
with $\alpha=5$, $a_\text{min}=\SI{1}{\micro m}$ and $a_\text{max}=\SI{6}{\micro m}$. With these input
parameters, the \emph{Ph\'enix} code handles the particle transport during the entire simulation, from
$t=\SI{0}{}$ to $t=\SI{27}{\micro s}$. This data is then given as input data in the Monte~Carlo simulation to
compute the corresponding expected spectrogram for the experimental setup characteristics. On the one hand,
for the experimental spectrogram, we represent
\begin{equation}
   P_\text{exp}(t,v)=10\log\left[\frac{K}{P_{0,\text{exp}}} S_\text{exp}\left(t,\frac{4\pi}{\lambda}v\right)\right]
\end{equation}
where $K$ is the global gain of the measuring setup and $P_{0,\text{exp}}$ is the
reference power. For this set of experiments, we have $K=\SI{e-6}{W^2V^{-2}s^{-2}}$ and choose $P_{0,\text{exp}}=\SI{1}{mW}$,
such that $P(t,v)$ is expressed in dBm. On the other hand, the simulated gain (in dB) is expressed as
\begin{equation}
   P_\text{sim}(t,v)=10\log\left[\frac{1}{T_w\left|\mathcal{A}_0\right|^2F_{0,\text{sim}}}S_\text{sim}\left(t,\frac{4\pi}{\lambda}v\right)\right]
\end{equation}
where the reference flux $F_{0,\text{sim}}=\SI{1}{V^{-2}s^{-1}}$. Even in the presence of the multiple
scattering regime, for both spectrograms frequencies are converted to \emph{apparent} velocities using the
single scattering relation $v=\omega/(4\pi)\lambda$. One has to keep in mind that the apparent velocity
correspond to the actual velocity of particles only in the single scattering regime. Given these definitions,
it is meaningful to compare the dynamic range rather than the absolute value.

\begin{figure}[!htb]
   \centering
   \includegraphics[width=0.8\linewidth]{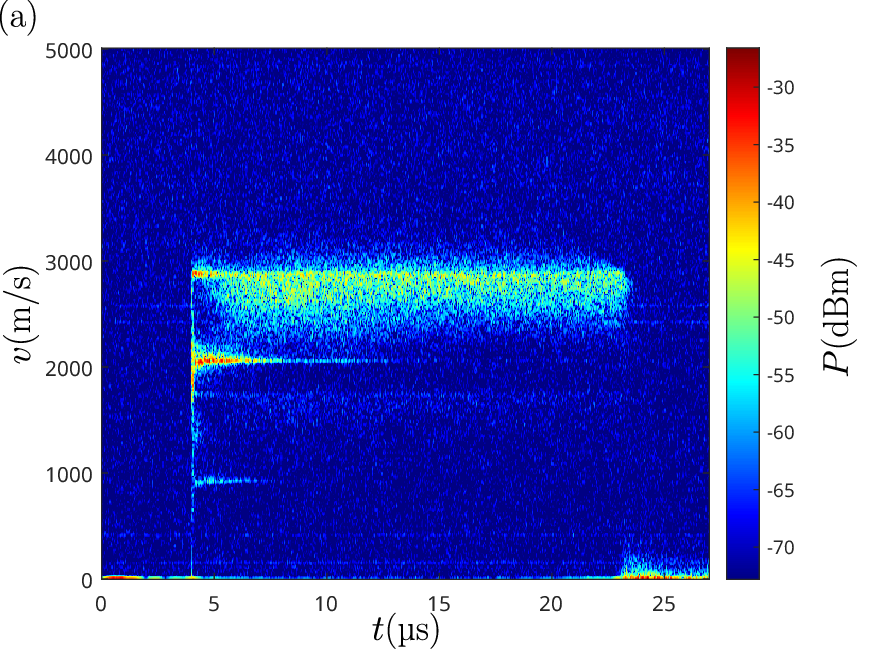}
   \includegraphics[width=0.8\linewidth]{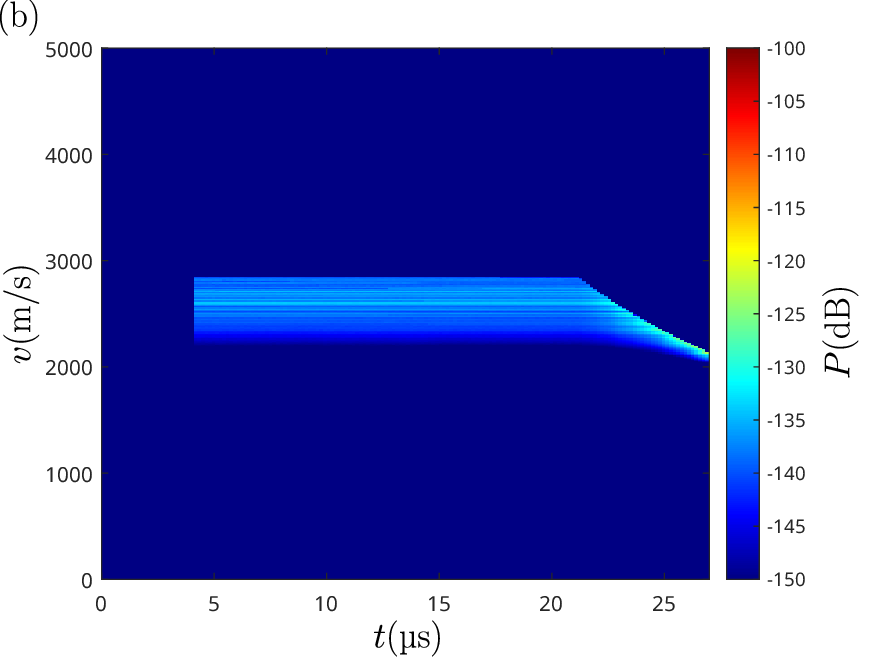}
   \includegraphics[width=0.8\linewidth]{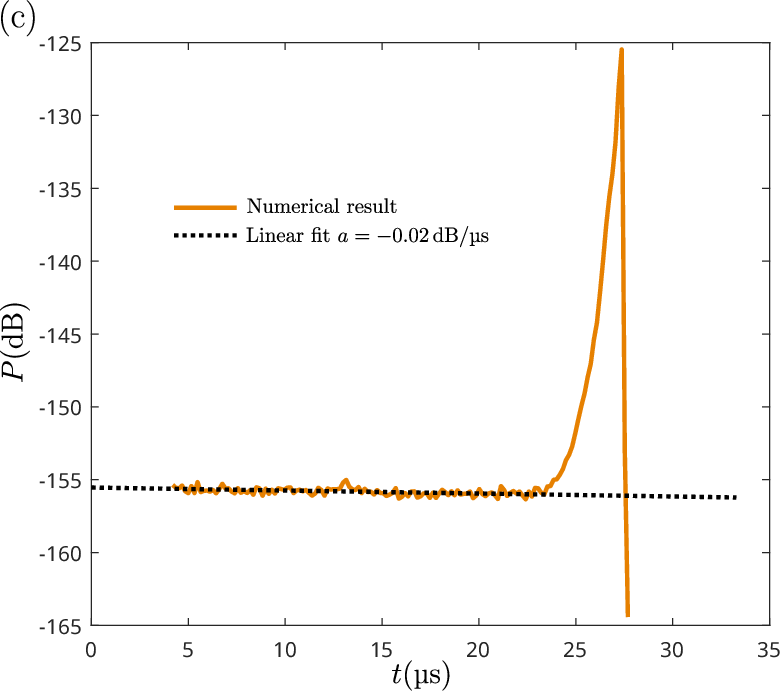}
   \caption{(a)~Experimental spectrogram in vacuum. The setup characteristics are given in
   Sec.~\ref{ejecta_creation}. The shock pressure is $P_\text{shock}=\SI{29.5}{GPa}$ and vacuum residual
   pressure was $P_\text{vacuum}=\SI{e-5}{bar}$. The ejecta is created at $t=\SI{4}{\micro s}$, it travels in
   ballistic expansion before reaching the probe at $t=\SI{23}{\micro s}$. (b)~Simulated spectrogram in vacuum
   with a power law size distribution of parameter $\alpha=5$. (c)~Extraction of the simulated spectrogram at
   $v=\SI{2100}{m/s}$ which illustrates the expected signal small decrease from $t=\SI{4}{}$ to
   $t=\SI{23}{\micro s}$.}
   \label{spectrogram_comparison_vacuum_power}
\end{figure}

Figure~\ref{spectrogram_comparison_vacuum_power} displays the comparison between the experimental spectrogram
in vacuum [Fig.~\ref{spectrogram_comparison_vacuum_power}(a)] and the first simulated spectrogram of this
study [Fig.~\ref{spectrogram_comparison_vacuum_power}(b)]. The first interesting observation is that, while
the levels differ, the dynamic ranges of both spectrograms are similar - around $\SI{50}{dB}$. Secondly, we
see that the dynamic of velocities as expected between $v=\SI{2000}{m/s}$ and $v=\SI{3000}{m/s}$ and that from
$t=\SI{4}{}$ to $t=\SI{23}{\micro s}$, the spectrogram does not depend much on time. We only see a
slight decrease on velocity readings as displayed for example for $v=\SI{2100}{m/s}$ in
Fig.~\ref{spectrogram_comparison_vacuum_power}(c). To understand this second observation, we have to consider
the behavior of particles in vacuum.

In vacuum, the size-velocity distribution of the particles remains constant due to ballistic transport. The
velocity difference between the front and the back particles will therefore stretch the ejecta along the
$z$-axis linearly over time. Since the size-velocity distribution does not change, this stretch does not
impact the local phase function but nonetheless it causes the particle number density to decrease as $1/t$ and
therefore the mean-free paths to increase linearly with $t$. To predict the impact on the specific intensity
measured by the PDV probe, it is instructive to stick to the random walk picture. On the one hand, since the
mean-free paths expand at the same rate as the medium, all the random walks will expand accordingly, \ie the
light propagation in the ejecta is homothetic with time. If we consider the transverse profile of the specific
intensity distribution at the front of the ejecta, it spreads and therefore decreases quadratically with time
$t$ in the $xy$-plan. On the other hand, considering its narrow aperture of $\SI{4.2}{\milli rad}$, the
collection from the PDV probe mostly happens in a cylinder of diameter $\phi_p$ which does not change with
time. The combination of both these phenomena results in a decrease with time of the PDV signal. In practice,
considering that the particle are mostly forward scattering, the light spread at the front of the ejecta
remains small and the decrease is subtile as seen in Fig.~\ref{spectrogram_comparison_vacuum_power}(c). While
obvious in the single scattering regime, this observation is here extended to multiple scattering. A formal
explanation of this phenomena based on the RTE is provided in App.~\ref{spectrogram_invariance}. Finally, the
fastest particles reach the probe at $t=\SI{23}{\micro s}$ in both the experimental and simulated
spectrograms. In the experimental case, this causes a particle accretion on the probe and an almost immediate
loss of return signal. In the simulation, we do not account for this effect. The particles are simply removed
for the medium and we eventually recover the free surface at $t=\SI{27}{\micro s}$.

\begin{figure}[!htb]
   \centering
   \includegraphics[width=0.8\linewidth]{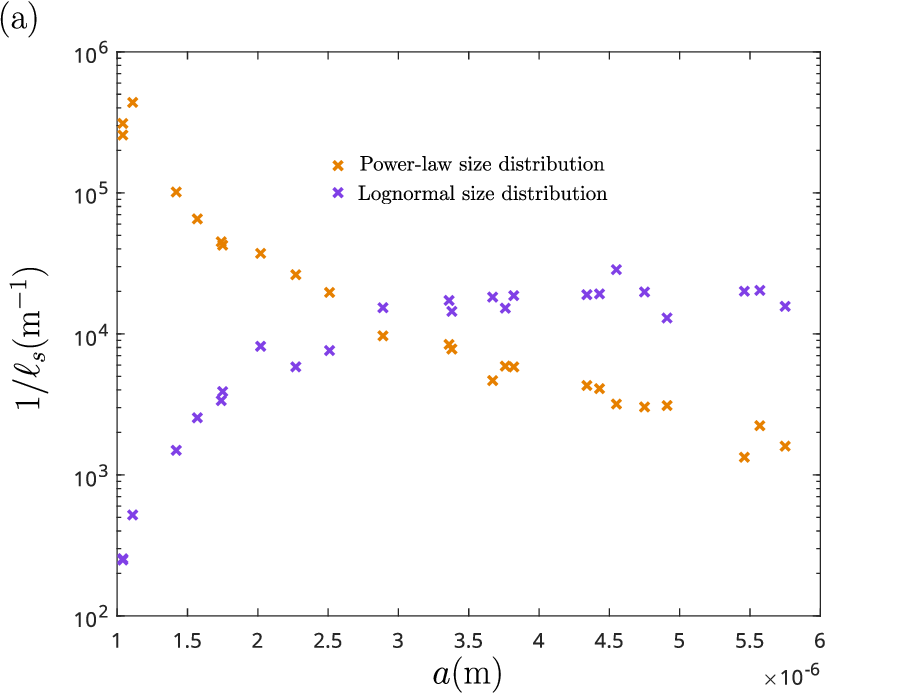}
   \includegraphics[width=0.8\linewidth]{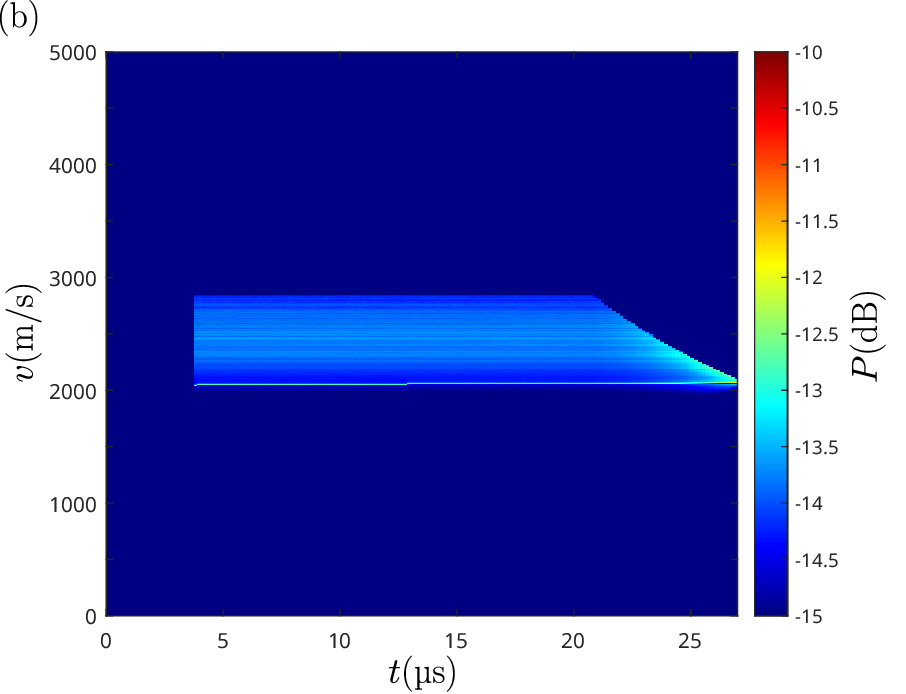}
   \caption{ (a)~Scattering mean-free path contributions over size for power law and lognormal initial
      particle size distribution.~(b)~Simulated spectrogram in vacuum with a lognormal size distribution of
      parameters $\sigma=0.5$ and $a_0=\SI{2.25}{\micro m}$.}
   \label{simulated_spectrogram_vacuum_lognormal}
\end{figure}

The key difference in the experimental spectrogram is that the free surface is visible almost over the full
duration of the experiment, while it does not appear in the simulation. This observation suggests that the
ejecta has a too large optical thickness $b$ for the free surface to be seen. Indeed, we find $b=16$ from
$t=\SI{4}{}$ to $t=\SI{23}{\micro s}$. To investigate this, we break down the scattering mean-free
path contributions of each numerical particle compared to its corresponding particle size, \ie we plot the
integrand of Eq.~(\ref{l_s}). We observe in Fig.~\ref{simulated_spectrogram_vacuum_lognormal}(a) that, while
they do not contribute much to the mass of the ejecta, a group of numerical particles with a small associated
size have the leading contribution to the scattering mean-free path. This ought to be the main reason why the
free surface remains hidden in the simulated spectrogram compared to the experimental one. This result argues
in favour of a size distribution with a lesser density at small particle sizes.

Beyond the standard power law distributions, lognormal size distributions have been proposed
lately~\cite{schauer_ejected_2017}. Since they tend to $0$ when $a$ tends to $0$, they precisely address the
divergence issue of power laws at small particle sizes. Therefore, we propose to change $h(a)$ to a lognormal
distribution which reads
\begin{equation}\label{lognormal}
   h(a)=
   \begin{cases}
      \frac{1}{a\sigma\sqrt{2\pi}}\exp\left[-\frac{\ln^2\left(a/a_0\right)}{2\sigma^2}\right], & \text{if}\,a\in\left[a_\text{min},a_\text{max}\right]\\
      0,&\text{otherwise}
   \end{cases}
\end{equation}
with $\sigma=0.5$, $a_0=\SI{2.25}{\micro m}$, $a_\text{min}=\SI{1}{\micro m}$ and $a_\text{max}=\SI{6}{\micro m}$.
In Fig.~\ref{simulated_spectrogram_vacuum_lognormal}(a), we see that for the same surface mass, the lognormal
distribution tends to attenuate the contribution of the numerical particles having a small associated size.
This results in a drastic decrease of the optical thickness to $b=6$ and should allow to recover the free
surface in the simulated spectrogram.

Figure~\ref{simulated_spectrogram_vacuum_lognormal}(b) represents the simulated spectrogram in vacuum for this
corrected size distribution. As expected, it allows one to recover the free surface in the spectrogram while
keeping all the already present and desired characteristics of the simulated spectrogram. A residual defect is
that free surface response seems to fade away quicker in the experimental spectrogram than in the simulated
one. We assume this effect is due the modelling of the free surface as a loss-less specular reflector instead
of properly accounting for the deformation induced by micro-jetting. This tends to overestimate its
contribution to the spectrogram.

This study of ejecta in vacuum is conclusive and gives us a first draft of ejecta description. Comparing
different simulated spectrograms to the experimental one allowed us to clearly favor one distribution over the
other only based on a PDV spectrogram. Before, a choice would have been harder to justify. The next step is a
more complex particle transport scenario - ejecta in helium - and the aim is to see if the description
established in vacuum holds.

\subsection{Drag coefficient effect study for ejecta in helium}\label{helium}

\begin{figure}[!htb]
   \centering
   \includegraphics[width=0.8\linewidth]{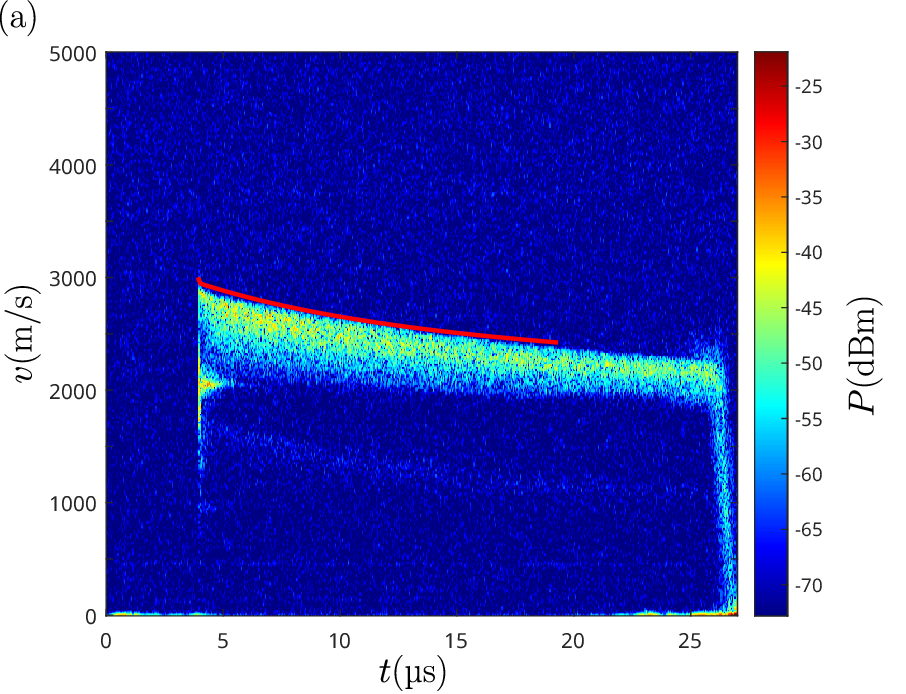}
   \includegraphics[width=0.8\linewidth]{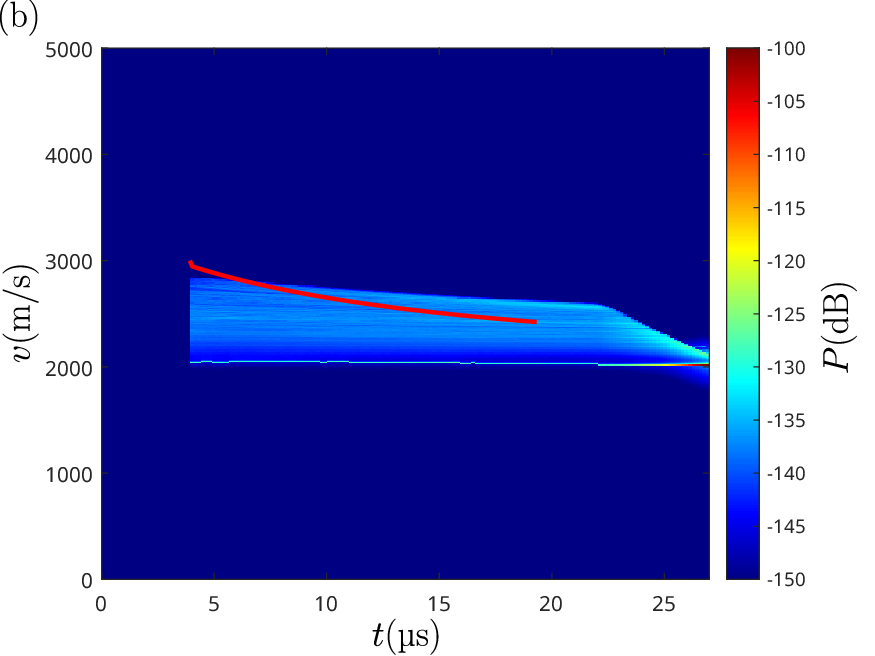}
   \includegraphics[width=0.8\linewidth]{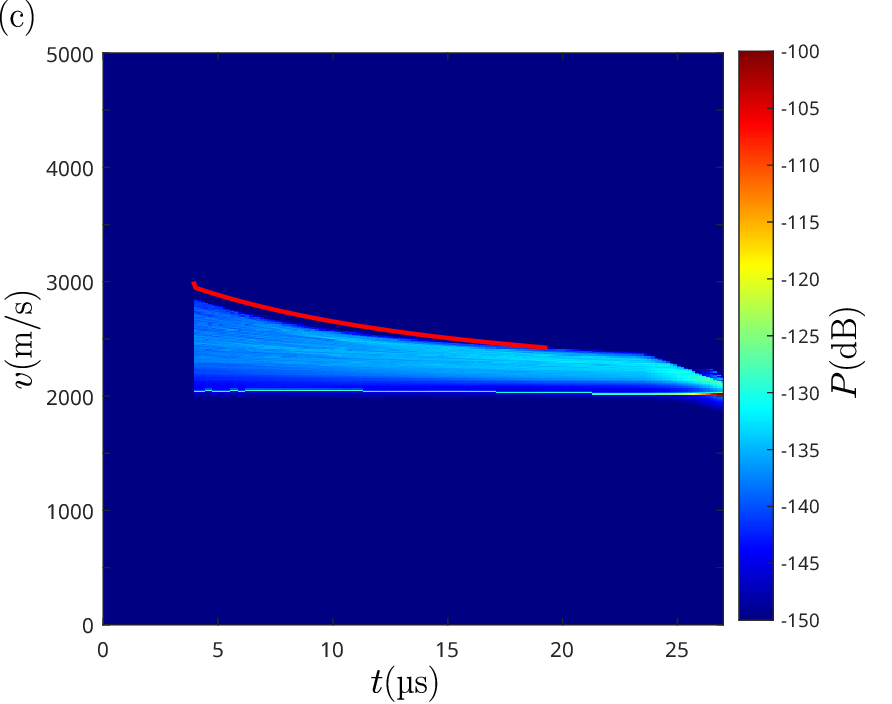}
   \caption{(a)~~Experimental spectrogram in helium. The setup characteristics are given in
   Sec.~\ref{ejecta_creation}. The shock pressure is $P_\text{shock}=\SI{29.5}{GPa}$ and helium pressure was
   $P_\text{helium}=\SI{5}{bar}$. The ejecta is created at $t=\SI{4}{\micro s}$, it travels in helium before
   reaching the probe at $t=\SI{26}{\micro s}$. The spectrogram is overlayed with an analytic fit of its upper
   boundary (red line). (b)~Simulated spectrogram in helium for a lognormal size distribution of parameters
   $\sigma=0.5$ and $a_0=\SI{2.25}{\micro m}$, overlayed with an analytic fit of the upper boundary of the
   experimental spectrogram (red line). (c)~Simulated spectrogram in helium for a lognormal size distribution
   of parameters $\sigma=0.5$ and $a_0=\SI{2.25}{\micro m}$, corrected drag force coefficients and overlayed with
   an analytic fit of the upper boundary of the experimental spectrogram (red line).}
   \label{spectrogram_comparison_helium}
\end{figure}

We now consider an ejecta in helium at $P_\text{helium}=\SI{5}{bar}$. We assume the same initial size distribution
as in vacuum, the lognormal distribution given in Eq.~(\ref{lognormal}), but now the presence of gas will
change the transport of particles as stated in Sec.\ref{ejecta_transport}. In case of helium, the shockwave in
gas, with a velocity $v=\SI{3084}{m/s}$, travels ahead of most of the ejecta. Therefore, particles travel in
shocked gas resulting in low particle-gas velocity differences. The Weber number $We_p$ given by
Eq.~(\ref{eq:weber}) remaining subcritical, the ejecta interacts with the gas mostly through drag forces.
Figure~\ref{spectrogram_comparison_helium}(a) represents the experimental spectrogram in helium. The
interaction with gas can be seen in the slowing down of the fastest particles, typically from $\SI{3000}{m/s}$
to $\SI{2500}{m/s}$. Since the drag force $\bm{F}_p$ scales with particle size as $a_p^2$, the slowing down of
particles scales in $a_p^{-1}$. Recovering the slowing down slop in simulated spectrograms would further
confirm the choice of the size distribution. To have a comparison point for the simulations, we compute a
numerical fit of the experimental spectrogram's upper boundary. It appears as a red overlay in
Fig.~\ref{spectrogram_comparison_helium}.

As for vacuum in Sec.~\ref{vacuum}, we use the \emph{Ph\'enix} code to compute the ejecta description
throughout transport in helium. We then use this output data to simulate the expected spectrogram in
Fig.~\ref{spectrogram_comparison_helium}(b) and compare it with the experimental one.
We see in Fig.\ref{spectrogram_comparison_helium}(b) that we kept the desired characteristics obtained in the
vacuum case, but this first spectrogram in helium does not fit the slowing down curve of the experimental
spectrogram. The slowing down of particles is underestimated, suggesting that the current drag force
$\bm{F}_p$ is undervalued. Considering the expression of $\bm{F}_p$ in Eq.~(\ref{eq:drag}), to increase drag
forces we can either shift back the particle size distribution towards smaller particles or increase the drag
coefficients $C_d$. Since we need to change the slowing down slope of the spectrogram while keeping the
current optical thickness, we have chosen to modify the drag coefficients $C_d$ to fit the experimental
spectrogram. After such a correction, we obtain another spectrogram in helium which is shown in
Fig.~\ref{spectrogram_comparison_helium}(c). This time, the slowing-down nicely fits the experimental upper
boundary. A down side remains the overvaluation of the free surface which remains visible in the simulated
spectrogram while it disappears at $t=\SI{8}{\micro s}$ in the experimental one. This bias of the model was
already discussed in Sec.~\ref{vacuum}. 

Now that we have a size distribution tested in the presence of a drag force, the next step is to see how it
holds up in a configuration with an additional interaction, \ie hydrodynamic break-up.

\subsection{Ejecta break-up model in air}\label{air}

\begin{figure}[!htb]
   \centering
   \includegraphics[width=0.8\linewidth]{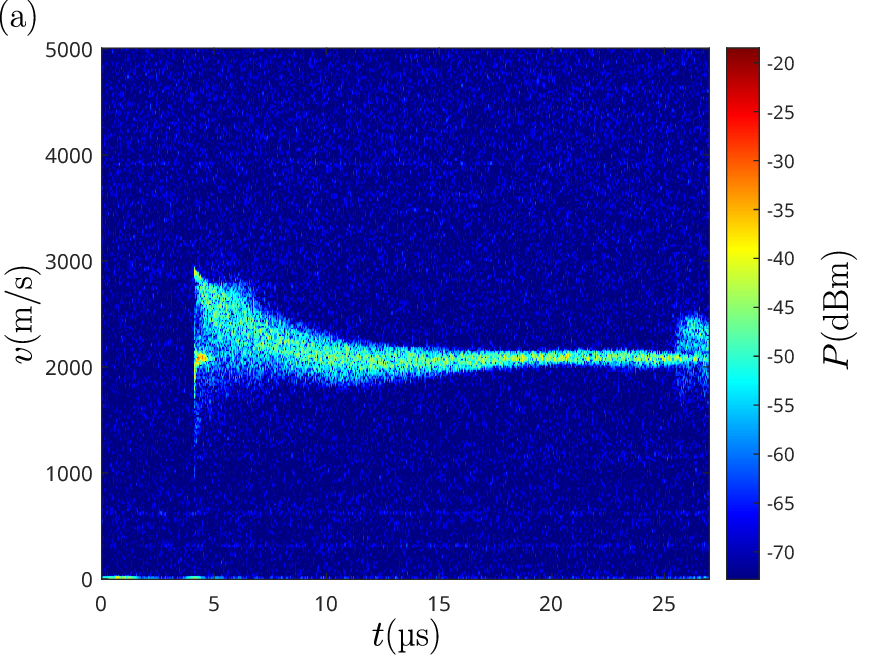}
   \includegraphics[width=0.8\linewidth]{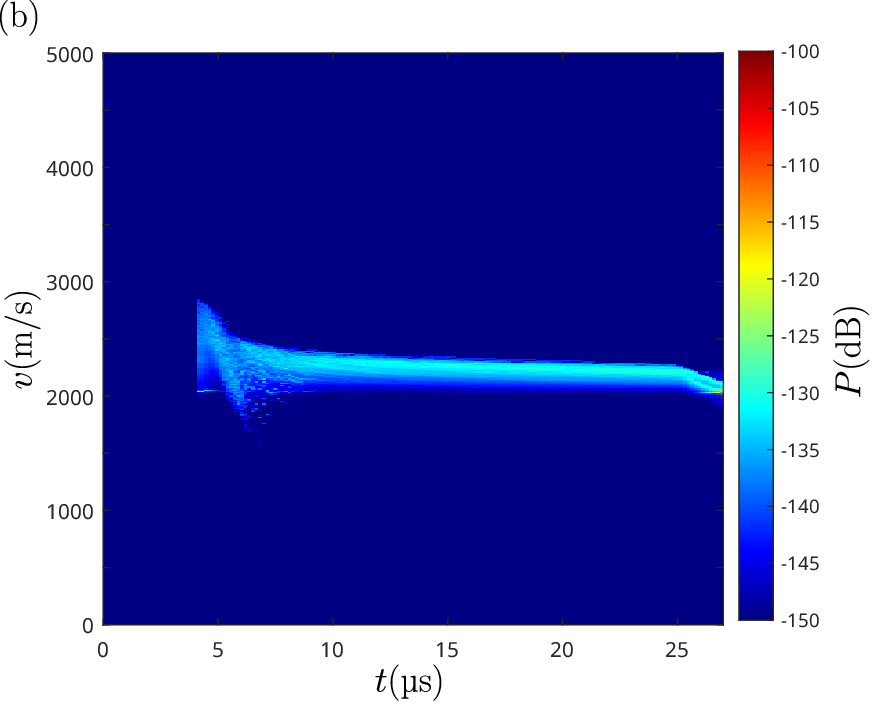}
   \caption{(a)~Experimental spectrogram in air. The setup characteristics are given in
   Sec.~\ref{ejecta_creation}. The shock pressure is $P_\text{shock}=\SI{29.5}{GPa}$ and air pressure is
   $P_\text{air}=\SI{1}{bar}$. The ejecta is created at $t=\SI{4}{\micro s}$. The ejecta front is heavily
   slown down and experience break-up in unshocked air until $t=\SI{10}{\micro s}$. From $t=\SI{10}{}$
   up to $t=\SI{20}{\micro s}$ the slowest particles fall back in shocked air and are reaccelerated. From
   $t=\SI{20}{}$ to $t=\SI{27}{\micro s}$, all particles seem at free surface velocity. (b)~Simulated
   spectrogram in air for a lognormal size distribution of parameters $\sigma=0.5$ and $a_0=\SI{2.25}{\micro m}$.}
   \label{spectrogram_comparison_air}
\end{figure}

We now consider the final and most complicated case, ejecta transport in air. We keep the same size
distribution as in helium, the lognormal distribution given by Eq.~(\ref{lognormal}), and check its relevance
in this new scenario. In air, the initial shockwave travels at $v=\SI{2520}{m/s}$. This means that while the
slowest particles travel in shocked air and, as in helium, interact with the gas mostly through drag forces,
the fastest particles travel in unshocked air. For the latter, if they are rather small, they slow down
rapidly through drag forces before being caught-up by the shockwave and eventually reaccelerated in shocked
air. If they are rather big, their Weber number $We_p$ is supcritical and they additionally experience
break-up. They may first slow down with a gentle slop but as soon as they break up, their reduced size makes
them slow down much faster before being caught by the shock wave and reaccelerated in the shocked air. In
Fig.~\ref{spectrogram_comparison_air}(a), displaying the experimental spectrogram in air, we can observe both
phenomena. Between $\SI{4}{}$ and $\SI{7}{\micro s}$, we can see a plateau around $\SI{2800}{m/s}$
corresponding to the fast particles before break-up. After break up, around $\SI{8}{\micro s}$, the free
surface gets screened and the slowing down is much more substantial. In the meantime, between $10$ and
$\SI{15}{\micro s}$, we see around $\SI{1800}{m/s}$ the reacceleration of the slowest particles. We aim to
capture these two phenomena in the simulated spectrogram.

Figure~\ref{spectrogram_comparison_air}(b) represents the simulated spectrogram in air. The expected key
features are clearly observed. Firstly, we observe that high velocity particles remain visible between
$t=\SI{4}{}$ and $t=\SI{6}{\micro s}$ before disappearing, which matches very well the experimental
spectrogram. Secondly, some particles are heavily slowed down and then reaccelerated by the shocked air
between $t=\SI{7}{}$ and $t=\SI{10}{\micro s}$. While this behavior is expected, it happens a bit too
early, we do not expect to see the reacceleration before $t=\SI{10}{\micro s}$. Thirdly, we keep the free
surface velocity in the early moment of the spectrogram before break-up screens it. Finally, the main issue is
the long term velocity distribution. While in the experiment, all velocities tend to the gas velocity of
$v=\SI{2060}{m/s}$, in the simulation a spreading between $v=\SI{2000}{m/s}$ and $v=\SI{2400}{m/s}$ remains.
We believe that the inability to perfectly match the experimental spectrogram puts forward the limit of
validity of the hypothesis made in Sec.~\ref{ejecta_creation} on the initial size-velocity distribution.
Namely, this observation argues in favor of a correlated size-velocity distribution. This question is
discussed in Sec.~\ref{discussion}.

\section{Discussion}\label{discussion}

A certain number of observations have been made in this study and it is worth discussing their implications,
and to compare them to results reported in the existing literature.

\subsection{Effect of particle break-up on the scattering mean-free path}\label{effect_break_up}

\begin{figure}[!htb]
   \centering
   \includegraphics[width=0.8\linewidth]{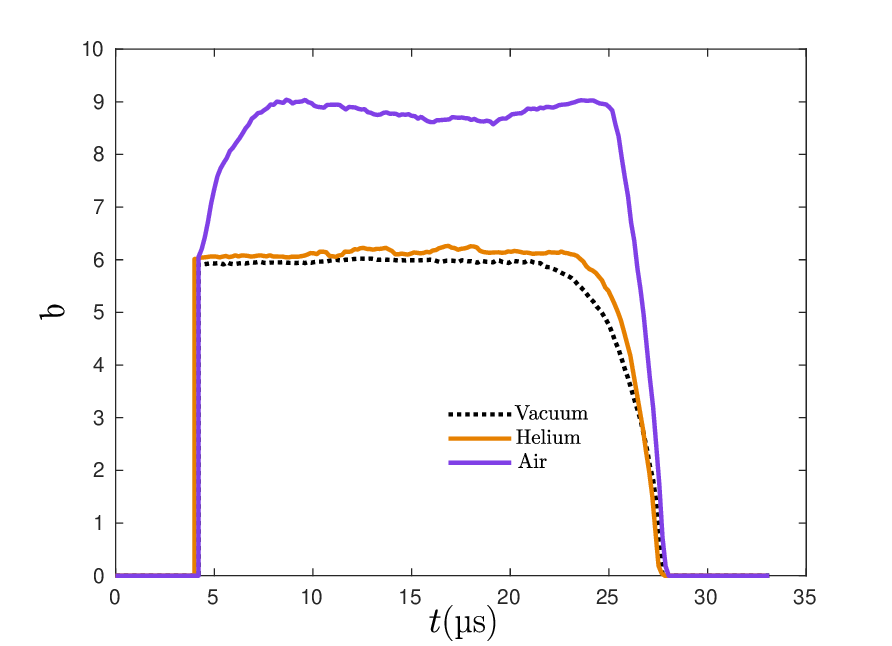}
   \caption{Optical thickness evolution during the simulation for each gas configuration. Using the simulation
   results from \emph{Ph\'enix} code and Eq.~(\ref{b}), the optical thickness is computed during the time
   corresponding to the experiments. The three ejecta simulations for a lognormal size distribution of
   parameters $\sigma=0.5$ and $a_0=\SI{2.25}{\micro m}$ seen in Sec.~\ref{results} are represented: vacuum
   (dotted black line), helium with corrected drag coefficients (solid orange line) and air (solid magenta
   line)}
   \label{optical_thickness_over_time}
\end{figure}

In Sec.~\ref{vacuum}, we showed that small particles between $\SI{1}{\micro m}$ and $\SI{3}{\micro m}$ had the
leading contribution to the scattering mean-free path. This observation was confirmed by the free surface
disappearance for ejecta in air in Sec.~\ref{air}. We attributed this phenomenon to an increase in the optical
thickness due to the break-up of initially large particles into numerous smaller ones. To check this
hypothesis, we have used the simulation results of the \emph{Ph\'enix} code and Eq.~(\ref{b}) to compute the
optical thickness during each simulation reported in Sec.~\ref{results}. The results are reported in
Fig.~\ref{optical_thickness_over_time}.
We see that for all three simulations, the initial optical thickness is around $b=6$, since all simulations
have the same initial size-velocity distribution. In the vacuum and helium cases, this optical thickness
remains constant until $t=\SI{23}{\micro s}$ where the ejecta gets shaved down from reaching the probe. It
eventually decreases back to zero by $t=\SI{27}{\micro s}$. In the case of air, we see that because of
fragmentation the optical thickness increases up to $b=9$ from $t=\SI{4}{}$ to $t=\SI{7}{\micro s}$. 

Buttler \emph{et al.}~\cite{buttler_understanding_2021} investigate ejecta in gas as well but consider a
reactive break-up scenario. In their study, a cerium ejecta is created and travels in deuterium gas. Through
an hydruration reaction between the metal and the gas, the initial particles also break up into multiple
smaller ones. This mechanism is similar to the hydrodynamic break-up in air we consider but interestingly
Buttler \emph{et al.} report opposite observations. In their experiment, they assume that the fragmentation of
the ejecta's front in smaller particles makes it invisible to the PDV wavelength, allowing to see particles
that up to that point had been hidden in the back of the ejecta. A key point to understand this difference is
to consider the size limit reached by each break-up mechanism. In the case of tin ejecta in shocked air we
reach an average size of $a=\SI{1}{\micro m}$ while Buttler \emph{et al.} assume the average size of particles
to be on the order of $a=\SI{100}{\nano m}$. For a mono-disperse ejecta of homogeneous particle number
density, the initial scattering mean-free path is $\ell_{s,0}=1/[\rho(a_0)\sigma_s(a_0)]$ with $a_0$ the
initial particle size and $\rho(a_0)$ the corresponding initial particle number density. If this ejecta were
to break up into smaller particles of size $a$ with mass conservation, the particle number density would
scales as $\rho(a)/\rho(a_0)=(a_0/a)^{3}$, leading to

\begin{equation}\label{l_s_break_up}
   \ell_s(a)=\frac{a^3}{a_0^3}\frac{1}{\rho(a_0)\sigma_s(a)}.
\end{equation}

\begin{figure}[!htb]
   \centering
   \includegraphics[width=0.8\linewidth]{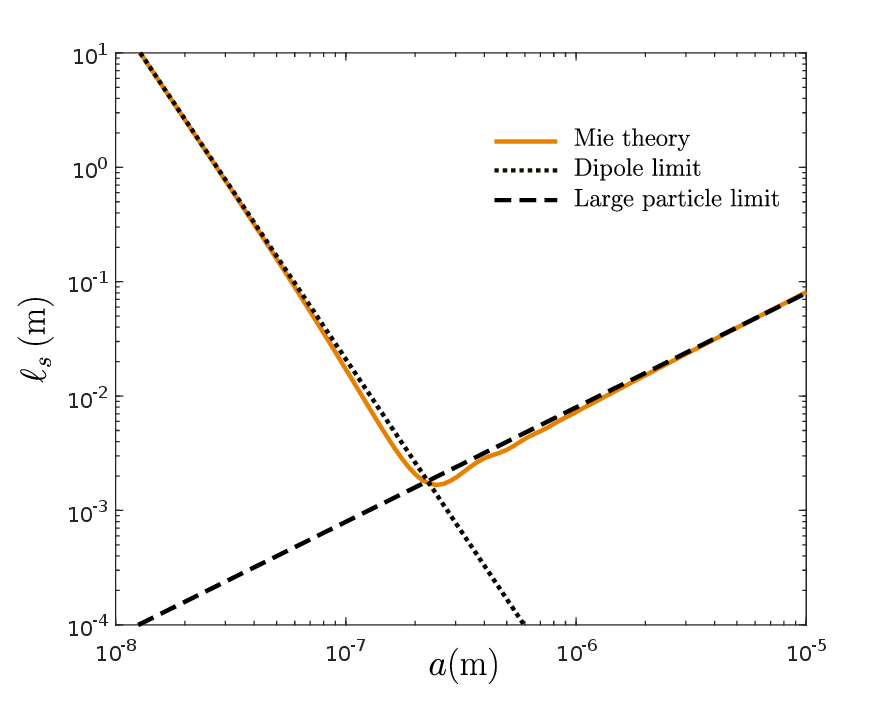}
   \caption{Scattering mean-free path for a size mono-disperse ejecta of tin in air experiencing
   fragmentation. The exact scattering mean-free paths is computed tanks to Mie theory (orange solid line).
   For comparison, we display the dipole limit case (black dotted line) and the large particle limit case
   (dashed solid). Mie theory allows one to describe the scattering mean-free path evolution over this large
   break-up window and recover limit cases at each end of it.}
   \label{l_s_break_up_fig}
\end{figure}

We have studied the scattering-mean free path given in Eq.~(\ref{l_s_break_up}) for a tin ejecta of an initial
size $a_0=\SI{10}{\micro m}$ and $\rho(a_0)=\SI{e14}{m^{-3}}$ and used Mie
theory~\cite{MIE-1908,schafer_matscat_2024} to compute the scattering cross-sections $\sigma_s$. In
Fig.~\ref{l_s_break_up_fig}, we report its variation from $\SI{10}{\micro m}$ down to $\SI{10}{\nano m}$. We
observe that, upon break-up and down to $a=\SI{250}{\nano m}$, the scattering mean-free path first decreases.
This is consistent with the phenomenon we observed for hydrodynamic break-up. It corresponds to the limit case
of particles much bigger than the wavelength. The scattering cross-section scales as $\sigma_s(a)\sim a^2$ and
therefore the scattering mean-free path as $\ell_s(a)\sim a$. Now, if we were to reach smaller particle sizes,
for example due to reactive break-up as in the study of Buttler \emph{et al.}, we see that the scattering
mean-free path would increase again - even exceeding its initial value below $a=\SI{60}{\nano m}$. We have
reached here the limit case of dipole approximation where the scattering cross-section scales as
$\sigma_s(a)\sim a^6$ and therefore the scattering mean-free path as $\ell_s(a)\sim a^{-3}$. Both limit cases
are represented in Fig.~\ref{l_s_break_up_fig}. We believe that this mechanism explains the phenomenon
observed by Buttler \emph{et al.} after multiple break-up cycles for cerium in deuterium. Break-up first
decreased the scattering mean-free path, as in our case, followed by an increase large enough to uncover
particles initially hidden.

The effects of reactive break-up were not opposite to the ones of hydrodynamic break-up, they were in fact
exceeding them. Going past the initial decrease in the scattering mean-free path, the increase for small
particle was enough to exceed the initial mean-free path.

\subsection{Introducing size and velocity dependencies}\label{size_velocity}

\begin{figure}[!htb]
   \centering
   \includegraphics[width=0.8\linewidth]{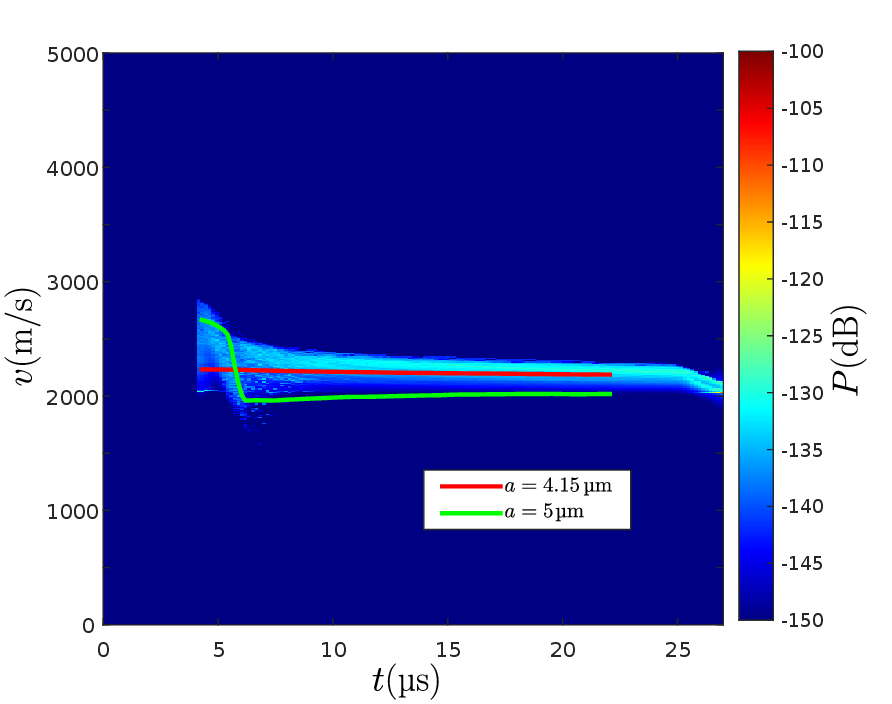}
   \caption{Simulated spectrogram in air for a lognormal size distribution of parameters $\sigma=0.5$ and
   $a_0=\SI{2.25}{\micro m}$. The simulated spectrogram is overlayed with the velocity curves of two interesting
   numerical particles extracted from the corresponding \emph{Ph\'enix} simulation: a particle of radius
   $a=\SI{4.15}{\micro m}$ with an initial velocity of $v=\SI{2234}{m/s}$ (red solid line) and a particle of
   radius $a=\SI{5}{\micro m}$ with an initial velocity of $v=\SI{2670}{m/s}$ (green solid line)}
   \label{spectrogram_discussion_air}
\end{figure}

As mentioned in Sec.~\ref{air}, the main issues in the simulated spectrogram show up at the end of the
experiment and in its early moments. At the end of the simulation, the spread in velocity is broader than the
one observed experimentally. We believe this is due to large particles, that sit right before the shockwave in
gas. These particles only travel in shocked air, therefore their velocity differential to the gas is low and
they do not break-up. In the meantime, they are too big to slow down only through drag force to the free
surface velocity and by the $\SI{20}{\micro s}$ mark. The velocity curve of such a particle with radius
$a=\SI{4.15}{\micro m}$ taken from the ejecta dynamics simulation is overlayed in red on the simulated
spectrogram in Fig.~\ref{spectrogram_discussion_air}. This observation argues in favor of a size velocity
distribution with bigger particles at low velocities and smaller particles at higher velocities. This is in
agreement with observations made by holographic measurements~\cite{sorenson_ejecta_2014} and the ejecta
mechanism suggested by molecular dynamics
simulations~\cite{durand_large-scale_2012,durand_power_2013,durand_mass-velocity_2015}.

Nonetheless, in the early moments, particles at the front of the ejecta must correspond to the high-velocity
plateau between between $t=\SI{4}{}$ and $t=\SI{6}{\micro s}$. These particles need to be big enough
so that despite of heavy drag forces in non shocked air they do not slow down immediately. When eventually
they break-up and slow down, they also allow to recover the reacceleration slope below the free surface
velocity between $t=\SI{7}{}$ and $t=\SI{10}{\micro s}$. A example of such a particle of radius
$a=\SI{5}{\micro m}$ taken from our simulation is overlayed in green on the simulated spectrogram in
Fig.~\ref{spectrogram_discussion_air}.

We believe this observation balances the previous one and illustrates the need of a more complex size velocity
distribution. Overall, we need big particles at the back of the ejecta and small particles at the front to
achieve the long term velocity profile. This distribution must then be completed with a few big particles
typically for the ejecta head, to observed the desired plateau in the early moments. While investigating such
correlations is beyond the scope of this article, this is an insightful observation. The ability to simulate
spectrogram has not only allowed to constrain size-velocity distributions to experimental spectrograms, but
also to confirm that the actual size velocity distributions of ejecta ought to be correlated distributions.

\section{Conclusion}

In summary, we have shown that PDV measurements could be used to retrieve additional information on the size
distribution of ejecta particles in shock compression experiments. Based on a exact relationship between the
specific intensity and the measured signal in PDV experiments and a rigorous RTE model for PDV experiments in
the multiple scattering regime, we have shown the influence of particle size distribution in PDV spectrograms.
To exploit this sensitivity, we implemented a simulation scheme allowing to use directly the results of ejecta
hydrodynamic simulations to compute simulated spectrograms. This opens up the possibility of indirect
size-velocity distribution evaluation thanks to different particle transport conditions accounting for drag
forces and particle break-up models. Finally, a comprehensive study on a real-conditions experiment showed
how, through an iterative process, spectrogram simulation allowed to better our ejecta description. We are not
aware of any technique that can describe in detail the partition of mass below the resolution limit of that
diagnostic. Indeed, we observed the effect of particles down to $a=\SI{60}{nm}$ while still working at
$\lambda=\SI{1.55}{\micro m}$.

From a more general point of view, this work is a proof of concept of a simulation chain aiming to mimic PDV
measurement in ejecta experiments. It shows that with a clear implementation of the direct problem including
the full path from an ejecta to its expected spectrogram, the comparison with experimental spectrogram already
allows one to have insightful ideas for the inverse problem of reconstructing this ejecta (or at least its
statistical properties). In the case of size-velocity distribution studies, this works argues in favor of
correlated size-velocity distributions, for the early moments of the experiment, corresponding to the
phenomenology captured by molecular dynamics simulations. The current simulation chain, including particle
transport and spectrogram simulation, would remain identical. Coupled with a better treatment of light
scattering close the free surface, typically around the micro-jets, we believe this would allow to recover
experimental spectrograms even in very complex cases as in air. This kind of indirect measurements could be
applied to many other ejecta scenarios, especially those with no analytical expression of the size-velocity
distribution throughout the experiment. For example, the transport of reactive ejecta in gas. In addition,
this opens the possibility of sensitivity studies on other parameters such as shock pressure, illumination
wavelength or material type. These are potential lines to be followed in further investigations.

\begin{acknowledgments}
   This work has received support under the program ``Investissements d’Avenir'' launched by the French
   Government. 
\end{acknowledgments}

\section*{Data Availability Statement}
The data that supports the findings of this study are available within the article.

\appendix
\section{Spectrogram invariance for single shock ejecta in vacuum}\label{spectrogram_invariance}

In the ejecta geometry where the mean-free path and the phase function depend only on depth, and in the
quasi-static approximation (slow cloud dynamics compared to the light travel time in the ejecta), the RTE
given in Eq.~(\ref{rte}) becomes
\begin{multline}\label{rte_ejecta}
   \left[\bm{u}\cdot\bm{\nabla}_{\bm{r}}+\frac{1}{\ell_e(\bm{r}\cdot\bm{u}_z,
   t,\omega)}\right]
   I(\bm{r},\bm{u},t,\omega)
\\
   =\frac{1}{\ell_s(\bm{r}\cdot\bm{u}_z,t,\omega)}\int p(\bm{r}\cdot\bm{u}_z,\bm{u},\bm{u}',t,\omega,\omega')
   I(\bm{r},\bm{u}',t,\omega')\ud\bm{u}'\frac{\ud \omega'}{2\pi}\,.
\end{multline}
Equation~(\ref{rte_ejecta}) taken at $\bm{r}\left(1+\Delta t/t\right)$ and $t+\Delta t$ reads
\begin{multline}\label{rte_homo}
   \left\{\bm{u}\cdot\bm{\nabla}_{\bm{r}}+\frac{1}{\ell_e\left[\bm{r}\left(1+\Delta t/t\right)\cdot\bm{u}_z,
   t+\Delta t,\omega\right]}\right\}
   I\left[\bm{r}\left(1+\Delta t/t\right),\bm{u},t+\Delta t,\omega\right]
\\
   =\frac{1}{\ell_s\left[\bm{r}\left(1+\Delta t/t\right)\cdot\bm{u}_z,t+\Delta t,\omega\right]}\int p\left[\bm{r}\left(1+\Delta t/t\right)\cdot\bm{u}_z,\bm{u},\bm{u}',t+\Delta t,\omega,\omega'\right]
\\\times
   I\left[\bm{r}\left(1+\Delta t/t\right),\bm{u}',t+\Delta t,\omega'\right]\ud\bm{u}'\frac{\ud \omega'}{2\pi}\, .
\end{multline}
In a single shock experiment in vacuum, the particles at the depth $\bm{r}\cdot{\bm{u}_z}$ at time $t$ are the
one that will be at the depth $\bm{r}\left(1+\Delta t/t\right)\cdot{\bm{u}_z}$ at time $t+\Delta t$. Therefore
the mean-free path and phase function obey the following conservation law
\begin{equation}
   p\left[ \bm{r}\left(1+\frac{\Delta t}{t}\right)\cdot{\bm{u}_z},\bm{u},\bm{u}',t+\Delta t,\omega,\omega'\right]
   =p\left(\bm{r}\cdot{\bm{u}_z},\bm{u},\bm{u}',t,\omega,\omega'\right),
\end{equation}
\begin{equation}
   \ell_{s,e}\left[\bm{r}\left(1+\frac{\Delta t}{t}\right)\cdot{\bm{u}_z},t+\Delta t\omega\right]
   =\ell_{s,e}\left(\bm{r}\cdot{\bm{u}_z},t,\omega\right)\left(1+\frac{\Delta t}{t}\right).
\end{equation}
While the phase function remains constant, for the mean-free paths the $\left(1+\Delta t/t\right)$ factor
accounts for the decrease in particle density caused by the homothetic stretch of the ejecta along the
$z$-axis. Using this property, Eq.~(\ref{rte_homo}) multiplied by $\left(1+\Delta t/t\right)$ reads
\begin{multline}\label{rte_homo_rescaled}
   \left[\bm{u}\cdot\bm{\nabla}_{\bm{r}}\left(1+\Delta t/t\right)+\frac{1}{\ell_e\left(\bm{r},t,\omega\right)}\right]
   I\left[\bm{r}\left(1+\Delta t/t\right),\bm{u},t+\Delta t,\omega\right]
\\
   =\frac{1}{\ell_s(\bm{r}\cdot\bm{u}_z,t,\omega)}\int p(\bm{r}\cdot\bm{u}_z,\bm{u},\bm{u}',t,\omega,\omega')
\\\times
   I\left[\bm{r}\left(1+\Delta t/t\right),\bm{u}',t+\Delta t,\omega'\right]\ud\bm{u}'\frac{\ud \omega'}{2\pi}\, .
\end{multline}
Comparing Eqs.~(\ref{rte_ejecta})~and~(\ref{rte_homo_rescaled}) shows that at $t$ and $t+\delta t$ the
specific intensity obeys to the same equation except that in the latter configuration all distances are to be
scaled up by a factor $1+\Delta t/t$, which formally proves the claim of Sec.~\ref{vacuum}.

%


\begin{thebibliography}{33}%
\makeatletter
\providecommand \@ifxundefined [1]{%
 \@ifx{#1\undefined}
}%
\providecommand \@ifnum [1]{%
 \ifnum #1\expandafter \@firstoftwo
 \else \expandafter \@secondoftwo
 \fi
}%
\providecommand \@ifx [1]{%
 \ifx #1\expandafter \@firstoftwo
 \else \expandafter \@secondoftwo
 \fi
}%
\providecommand \natexlab [1]{#1}%
\providecommand \enquote  [1]{``#1''}%
\providecommand \bibnamefont  [1]{#1}%
\providecommand \bibfnamefont [1]{#1}%
\providecommand \citenamefont [1]{#1}%
\providecommand \href@noop [0]{\@secondoftwo}%
\providecommand \href [0]{\begingroup \@sanitize@url \@href}%
\providecommand \@href[1]{\@@startlink{#1}\@@href}%
\providecommand \@@href[1]{\endgroup#1\@@endlink}%
\providecommand \@sanitize@url [0]{\catcode `\\12\catcode `\$12\catcode
  `\&12\catcode `\#12\catcode `\^12\catcode `\_12\catcode `\%12\relax}%
\providecommand \@@startlink[1]{}%
\providecommand \@@endlink[0]{}%
\providecommand \url  [0]{\begingroup\@sanitize@url \@url }%
\providecommand \@url [1]{\endgroup\@href {#1}{\urlprefix }}%
\providecommand \urlprefix  [0]{URL }%
\providecommand \Eprint [0]{\href }%
\providecommand \doibase [0]{http://dx.doi.org/}%
\providecommand \selectlanguage [0]{\@gobble}%
\providecommand \bibinfo  [0]{\@secondoftwo}%
\providecommand \bibfield  [0]{\@secondoftwo}%
\providecommand \translation [1]{[#1]}%
\providecommand \BibitemOpen [0]{}%
\providecommand \bibitemStop [0]{}%
\providecommand \bibitemNoStop [0]{.\EOS\space}%
\providecommand \EOS [0]{\spacefactor3000\relax}%
\providecommand \BibitemShut  [1]{\csname bibitem#1\endcsname}%
\let\auto@bib@innerbib\@empty
\bibitem [{\citenamefont {Buttler}\ \emph {et~al.}(2017)\citenamefont
  {Buttler}, \citenamefont {Williams},\ and\ \citenamefont
  {Najjar}}]{buttler_foreword_2017}%
  \BibitemOpen
  \bibfield  {author} {\bibinfo {author} {\bibfnamefont {W.~T.}\ \bibnamefont
  {Buttler}}, \bibinfo {author} {\bibfnamefont {R.~J.~R.}\ \bibnamefont
  {Williams}}, \ and\ \bibinfo {author} {\bibfnamefont {F.~M.}\ \bibnamefont
  {Najjar}},\ }\href {\doibase 10.1007/s40870-017-0120-8} {\bibfield  {journal}
  {\bibinfo  {journal} {J. Dyn. Behav. Mater.}\ }\textbf {\bibinfo {volume}
  {3}},\ \bibinfo {pages} {151} (\bibinfo {year} {2017})}\BibitemShut {NoStop}%
\bibitem [{\citenamefont {Richtmyer}(1960)}]{richtmyer_taylor_1960}%
  \BibitemOpen
  \bibfield  {author} {\bibinfo {author} {\bibfnamefont {R.~D.}\ \bibnamefont
  {Richtmyer}},\ }\href {\doibase 10.1002/cpa.3160130207} {\bibfield  {journal}
  {\bibinfo  {journal} {Comm Pure Appl Math}\ }\textbf {\bibinfo {volume}
  {13}},\ \bibinfo {pages} {297} (\bibinfo {year} {1960})}\BibitemShut
  {NoStop}%
\bibitem [{\citenamefont {Meshkov}(1972)}]{meshkov_instability_1972}%
  \BibitemOpen
  \bibfield  {author} {\bibinfo {author} {\bibfnamefont {E.~E.}\ \bibnamefont
  {Meshkov}},\ }\href {\doibase 10.1007/BF01015969} {\bibfield  {journal}
  {\bibinfo  {journal} {Fluid Dyn}\ }\textbf {\bibinfo {volume} {4}},\ \bibinfo
  {pages} {101} (\bibinfo {year} {1972})}\BibitemShut {NoStop}%
\bibitem [{\citenamefont {Asay}\ \emph {et~al.}(1976)\citenamefont {Asay},
  \citenamefont {Mix},\ and\ \citenamefont {Perry}}]{asay_ejection_1976}%
  \BibitemOpen
  \bibfield  {author} {\bibinfo {author} {\bibfnamefont {J.~R.}\ \bibnamefont
  {Asay}}, \bibinfo {author} {\bibfnamefont {L.~P.}\ \bibnamefont {Mix}}, \
  and\ \bibinfo {author} {\bibfnamefont {F.~C.}\ \bibnamefont {Perry}},\ }\href
  {\doibase 10.1063/1.89066} {\bibfield  {journal} {\bibinfo  {journal} {Appl.
  Phys. Lett.}\ }\textbf {\bibinfo {volume} {29}},\ \bibinfo {pages} {284}
  (\bibinfo {year} {1976})}\BibitemShut {NoStop}%
\bibitem [{\citenamefont {Andriot}\ \emph {et~al.}(1982)\citenamefont
  {Andriot}, \citenamefont {Chapron},\ and\ \citenamefont
  {Olive}}]{andriot_ejection_1982}%
  \BibitemOpen
  \bibfield  {author} {\bibinfo {author} {\bibfnamefont {P.}~\bibnamefont
  {Andriot}}, \bibinfo {author} {\bibfnamefont {P.}~\bibnamefont {Chapron}}, \
  and\ \bibinfo {author} {\bibfnamefont {F.}~\bibnamefont {Olive}},\ }in\ \href
  {\doibase 10.1063/1.33316} {\emph {\bibinfo {booktitle} {{AIP} {Conference}
  {Proceeding} {Volume} 78}}}\ (\bibinfo  {publisher} {AIP},\ \bibinfo {year}
  {1982})\ pp.\ \bibinfo {pages} {505--509}\BibitemShut {NoStop}%
\bibitem [{\citenamefont {Buttler}\ \emph {et~al.}(2012)\citenamefont
  {Buttler}, \citenamefont {Oró}, \citenamefont {Preston}, \citenamefont
  {Mikaelian}, \citenamefont {Cherne}, \citenamefont {Hixson}, \citenamefont
  {Mariam}, \citenamefont {Morris}, \citenamefont {Stone}, \citenamefont
  {Terrones},\ and\ \citenamefont {Tupa}}]{buttler_unstable_2012}%
  \BibitemOpen
  \bibfield  {author} {\bibinfo {author} {\bibfnamefont {W.~T.}\ \bibnamefont
  {Buttler}}, \bibinfo {author} {\bibfnamefont {D.~M.}\ \bibnamefont {Oró}},
  \bibinfo {author} {\bibfnamefont {D.~L.}\ \bibnamefont {Preston}}, \bibinfo
  {author} {\bibfnamefont {K.~O.}\ \bibnamefont {Mikaelian}}, \bibinfo {author}
  {\bibfnamefont {F.~J.}\ \bibnamefont {Cherne}}, \bibinfo {author}
  {\bibfnamefont {R.~S.}\ \bibnamefont {Hixson}}, \bibinfo {author}
  {\bibfnamefont {F.~G.}\ \bibnamefont {Mariam}}, \bibinfo {author}
  {\bibfnamefont {C.}~\bibnamefont {Morris}}, \bibinfo {author} {\bibfnamefont
  {J.~B.}\ \bibnamefont {Stone}}, \bibinfo {author} {\bibfnamefont
  {G.}~\bibnamefont {Terrones}}, \ and\ \bibinfo {author} {\bibfnamefont
  {D.}~\bibnamefont {Tupa}},\ }\href {\doibase 10.1017/jfm.2012.190} {\bibfield
   {journal} {\bibinfo  {journal} {J. Fluid Mech.}\ }\textbf {\bibinfo {volume}
  {703}},\ \bibinfo {pages} {60} (\bibinfo {year} {2012})}\BibitemShut
  {NoStop}%
\bibitem [{\citenamefont {Dimonte}\ \emph {et~al.}(2013)\citenamefont
  {Dimonte}, \citenamefont {Terrones}, \citenamefont {Cherne},\ and\
  \citenamefont {Ramaprabhu}}]{dimonte_ejecta_2013}%
  \BibitemOpen
  \bibfield  {author} {\bibinfo {author} {\bibfnamefont {G.}~\bibnamefont
  {Dimonte}}, \bibinfo {author} {\bibfnamefont {G.}~\bibnamefont {Terrones}},
  \bibinfo {author} {\bibfnamefont {F.~J.}\ \bibnamefont {Cherne}}, \ and\
  \bibinfo {author} {\bibfnamefont {P.}~\bibnamefont {Ramaprabhu}},\ }\href
  {\doibase 10.1063/1.4773575} {\bibfield  {journal} {\bibinfo  {journal} {J.
  Appl. Phys.}\ }\textbf {\bibinfo {volume} {113}},\ \bibinfo {pages} {024905}
  (\bibinfo {year} {2013})}\BibitemShut {NoStop}%
\bibitem [{\citenamefont {Georgievskaya}\ and\ \citenamefont
  {Raevsky}(2017)}]{georgievskaya_model_2017}%
  \BibitemOpen
  \bibfield  {author} {\bibinfo {author} {\bibfnamefont {A.~B.}\ \bibnamefont
  {Georgievskaya}}\ and\ \bibinfo {author} {\bibfnamefont {V.~A.}\ \bibnamefont
  {Raevsky}},\ }\href {\doibase 10.1007/s40870-017-0118-2} {\bibfield
  {journal} {\bibinfo  {journal} {J. dynamic behavior mater.}\ }\textbf
  {\bibinfo {volume} {3}},\ \bibinfo {pages} {321} (\bibinfo {year}
  {2017})}\BibitemShut {NoStop}%
\bibitem [{\citenamefont {Fung}\ \emph {et~al.}(2013)\citenamefont {Fung},
  \citenamefont {Harrison}, \citenamefont {Chitanvis},\ and\ \citenamefont
  {Margulies}}]{fung_ejecta_2013}%
  \BibitemOpen
  \bibfield  {author} {\bibinfo {author} {\bibfnamefont {J.}~\bibnamefont
  {Fung}}, \bibinfo {author} {\bibfnamefont {A.}~\bibnamefont {Harrison}},
  \bibinfo {author} {\bibfnamefont {S.}~\bibnamefont {Chitanvis}}, \ and\
  \bibinfo {author} {\bibfnamefont {J.}~\bibnamefont {Margulies}},\ }\href
  {\doibase 10.1016/j.compfluid.2012.08.011} {\bibfield  {journal} {\bibinfo
  {journal} {Comput. Fluids}\ }\textbf {\bibinfo {volume} {83}},\ \bibinfo
  {pages} {177} (\bibinfo {year} {2013})}\BibitemShut {NoStop}%
\bibitem [{\citenamefont {Durand}\ and\ \citenamefont
  {Soulard}(2012)}]{durand_large-scale_2012}%
  \BibitemOpen
  \bibfield  {author} {\bibinfo {author} {\bibfnamefont {O.}~\bibnamefont
  {Durand}}\ and\ \bibinfo {author} {\bibfnamefont {L.}~\bibnamefont
  {Soulard}},\ }\href {\doibase 10.1063/1.3684978} {\bibfield  {journal}
  {\bibinfo  {journal} {J. Appl. Phys.}\ }\textbf {\bibinfo {volume} {111}},\
  \bibinfo {pages} {044901} (\bibinfo {year} {2012})}\BibitemShut {NoStop}%
\bibitem [{\citenamefont {Durand}\ and\ \citenamefont
  {Soulard}(2013)}]{durand_power_2013}%
  \BibitemOpen
  \bibfield  {author} {\bibinfo {author} {\bibfnamefont {O.}~\bibnamefont
  {Durand}}\ and\ \bibinfo {author} {\bibfnamefont {L.}~\bibnamefont
  {Soulard}},\ }\href {\doibase 10.1063/1.4832758} {\bibfield  {journal}
  {\bibinfo  {journal} {J. Appl. Phys.}\ }\textbf {\bibinfo {volume} {114}},\
  \bibinfo {pages} {194902} (\bibinfo {year} {2013})}\BibitemShut {NoStop}%
\bibitem [{\citenamefont {Durand}\ and\ \citenamefont
  {Soulard}(2015)}]{durand_mass-velocity_2015}%
  \BibitemOpen
  \bibfield  {author} {\bibinfo {author} {\bibfnamefont {O.}~\bibnamefont
  {Durand}}\ and\ \bibinfo {author} {\bibfnamefont {L.}~\bibnamefont
  {Soulard}},\ }\href {\doibase 10.1063/1.4918537} {\bibfield  {journal}
  {\bibinfo  {journal} {J. Appl. Phys.}\ }\textbf {\bibinfo {volume} {117}},\
  \bibinfo {pages} {165903} (\bibinfo {year} {2015})}\BibitemShut {NoStop}%
\bibitem [{\citenamefont {Monfared}\ \emph {et~al.}(2015)\citenamefont
  {Monfared}, \citenamefont {Buttler}, \citenamefont {Frayer}, \citenamefont
  {Grover}, \citenamefont {LaLone}, \citenamefont {Stevens}, \citenamefont
  {Stone}, \citenamefont {Turley},\ and\ \citenamefont
  {Schauer}}]{monfared_ejected_2015}%
  \BibitemOpen
  \bibfield  {author} {\bibinfo {author} {\bibfnamefont {S.~K.}\ \bibnamefont
  {Monfared}}, \bibinfo {author} {\bibfnamefont {W.~T.}\ \bibnamefont
  {Buttler}}, \bibinfo {author} {\bibfnamefont {D.~K.}\ \bibnamefont {Frayer}},
  \bibinfo {author} {\bibfnamefont {M.}~\bibnamefont {Grover}}, \bibinfo
  {author} {\bibfnamefont {B.~M.}\ \bibnamefont {LaLone}}, \bibinfo {author}
  {\bibfnamefont {G.~D.}\ \bibnamefont {Stevens}}, \bibinfo {author}
  {\bibfnamefont {J.~B.}\ \bibnamefont {Stone}}, \bibinfo {author}
  {\bibfnamefont {W.~D.}\ \bibnamefont {Turley}}, \ and\ \bibinfo {author}
  {\bibfnamefont {M.~M.}\ \bibnamefont {Schauer}},\ }\href {\doibase
  10.1063/1.4922180} {\bibfield  {journal} {\bibinfo  {journal} {Journal of
  Applied Physics}\ }\textbf {\bibinfo {volume} {117}},\ \bibinfo {pages}
  {223105} (\bibinfo {year} {2015})}\BibitemShut {NoStop}%
\bibitem [{\citenamefont {Schauer}\ \emph {et~al.}(2017)\citenamefont
  {Schauer}, \citenamefont {Buttler}, \citenamefont {Frayer}, \citenamefont
  {Grover}, \citenamefont {LaLone}, \citenamefont {Monfared}, \citenamefont
  {Sorenson}, \citenamefont {Stevens},\ and\ \citenamefont
  {Turley}}]{schauer_ejected_2017}%
  \BibitemOpen
  \bibfield  {author} {\bibinfo {author} {\bibfnamefont {M.~M.}\ \bibnamefont
  {Schauer}}, \bibinfo {author} {\bibfnamefont {W.~T.}\ \bibnamefont
  {Buttler}}, \bibinfo {author} {\bibfnamefont {D.~K.}\ \bibnamefont {Frayer}},
  \bibinfo {author} {\bibfnamefont {M.}~\bibnamefont {Grover}}, \bibinfo
  {author} {\bibfnamefont {B.~M.}\ \bibnamefont {LaLone}}, \bibinfo {author}
  {\bibfnamefont {S.~K.}\ \bibnamefont {Monfared}}, \bibinfo {author}
  {\bibfnamefont {D.~S.}\ \bibnamefont {Sorenson}}, \bibinfo {author}
  {\bibfnamefont {G.~D.}\ \bibnamefont {Stevens}}, \ and\ \bibinfo {author}
  {\bibfnamefont {W.~D.}\ \bibnamefont {Turley}},\ }\href {\doibase
  10.1007/s40870-017-0111-9} {\bibfield  {journal} {\bibinfo  {journal} {J.
  Dyn. Behav. Mater.}\ }\textbf {\bibinfo {volume} {3}} (\bibinfo {year}
  {2017}),\ 10.1007/s40870-017-0111-9}\BibitemShut {NoStop}%
\bibitem [{\citenamefont {Sorenson}\ \emph {et~al.}(2014)\citenamefont
  {Sorenson}, \citenamefont {Pazuchanics}, \citenamefont {Johnson},
  \citenamefont {Malone}, \citenamefont {Kaufman}, \citenamefont {Tibbitts},
  \citenamefont {Tunnell}, \citenamefont {Marks}, \citenamefont {Capelle},
  \citenamefont {Grover}, \citenamefont {Marshall}, \citenamefont {Stevens},
  \citenamefont {Turley},\ and\ \citenamefont {LaLone}}]{sorenson_ejecta_2014}%
  \BibitemOpen
  \bibfield  {author} {\bibinfo {author} {\bibfnamefont {D.~S.}\ \bibnamefont
  {Sorenson}}, \bibinfo {author} {\bibfnamefont {P.}~\bibnamefont
  {Pazuchanics}}, \bibinfo {author} {\bibfnamefont {R.}~\bibnamefont
  {Johnson}}, \bibinfo {author} {\bibfnamefont {R.~M.}\ \bibnamefont {Malone}},
  \bibinfo {author} {\bibfnamefont {M.~I.}\ \bibnamefont {Kaufman}}, \bibinfo
  {author} {\bibfnamefont {A.}~\bibnamefont {Tibbitts}}, \bibinfo {author}
  {\bibfnamefont {T.}~\bibnamefont {Tunnell}}, \bibinfo {author} {\bibfnamefont
  {D.}~\bibnamefont {Marks}}, \bibinfo {author} {\bibfnamefont {G.~A.}\
  \bibnamefont {Capelle}}, \bibinfo {author} {\bibfnamefont {M.}~\bibnamefont
  {Grover}}, \bibinfo {author} {\bibfnamefont {B.}~\bibnamefont {Marshall}},
  \bibinfo {author} {\bibfnamefont {G.~D.}\ \bibnamefont {Stevens}}, \bibinfo
  {author} {\bibfnamefont {W.~D.}\ \bibnamefont {Turley}}, \ and\ \bibinfo
  {author} {\bibfnamefont {B.}~\bibnamefont {LaLone}},\ }\href@noop {} {\emph
  {\bibinfo {title} {Ejecta {Particle}-{Size} {Measurements} in {Vacuum} and
  {Helium} {Gas} using {Ultraviolet} {In}-{Line} {Fraunhofer} {Holography}}}},\
  \bibinfo {type} {Tech. Rep.}\ \bibinfo {number} {LA-UR-14-24722}\ (\bibinfo
  {institution} {National Nuclear Security Administration},\ \bibinfo {year}
  {2014})\BibitemShut {NoStop}%
\bibitem [{\citenamefont {Guildenbecher}\ \emph {et~al.}(2023)\citenamefont
  {Guildenbecher}, \citenamefont {McMaster}, \citenamefont {Corredor},
  \citenamefont {Malone}, \citenamefont {Mance}, \citenamefont {Rudziensky},
  \citenamefont {Sorenson}, \citenamefont {Danielson},\ and\ \citenamefont
  {Duke}}]{guildenbecher_ultraviolet_2023}%
  \BibitemOpen
  \bibfield  {author} {\bibinfo {author} {\bibfnamefont {D.~R.}\ \bibnamefont
  {Guildenbecher}}, \bibinfo {author} {\bibfnamefont {A.}~\bibnamefont
  {McMaster}}, \bibinfo {author} {\bibfnamefont {A.}~\bibnamefont {Corredor}},
  \bibinfo {author} {\bibfnamefont {B.}~\bibnamefont {Malone}}, \bibinfo
  {author} {\bibfnamefont {J.}~\bibnamefont {Mance}}, \bibinfo {author}
  {\bibfnamefont {E.}~\bibnamefont {Rudziensky}}, \bibinfo {author}
  {\bibfnamefont {D.}~\bibnamefont {Sorenson}}, \bibinfo {author}
  {\bibfnamefont {J.}~\bibnamefont {Danielson}}, \ and\ \bibinfo {author}
  {\bibfnamefont {D.~L.}\ \bibnamefont {Duke}},\ }\href {\doibase
  10.1364/OE.486461} {\bibfield  {journal} {\bibinfo  {journal} {Opt. Express}\
  }\textbf {\bibinfo {volume} {31}},\ \bibinfo {pages} {14911} (\bibinfo {year}
  {2023})}\BibitemShut {NoStop}%
\bibitem [{\citenamefont {{S}trand}\ \emph {et~al.}(2006)\citenamefont
  {{S}trand}, \citenamefont {{G}oosman}, \citenamefont {{M}artinez},
  \citenamefont {{W}hitworth},\ and\ \citenamefont {{K}uhlow}}]{STRAND-2006}%
  \BibitemOpen
  \bibfield  {author} {\bibinfo {author} {\bibfnamefont {O.~T.}\ \bibnamefont
  {{S}trand}}, \bibinfo {author} {\bibfnamefont {D.~R.}\ \bibnamefont
  {{G}oosman}}, \bibinfo {author} {\bibfnamefont {C.}~\bibnamefont
  {{M}artinez}}, \bibinfo {author} {\bibfnamefont {T.~L.}\ \bibnamefont
  {{W}hitworth}}, \ and\ \bibinfo {author} {\bibfnamefont {W.~W.}\ \bibnamefont
  {{K}uhlow}},\ }\href {\doibase 10.1063/1.2336749} {\bibfield  {journal}
  {\bibinfo  {journal} {Rev. Sci. Instrum.}\ }\textbf {\bibinfo {volume}
  {77}},\ \bibinfo {pages} {083108} (\bibinfo {year} {2006})}\BibitemShut
  {NoStop}%
\bibitem [{\citenamefont {Mercier}\ \emph {et~al.}(2006)\citenamefont
  {Mercier}, \citenamefont {Benier}, \citenamefont {Azzolina}, \citenamefont
  {Lagrange},\ and\ \citenamefont {Partouche}}]{mercier_photonic_2006}%
  \BibitemOpen
  \bibfield  {author} {\bibinfo {author} {\bibfnamefont {P.}~\bibnamefont
  {Mercier}}, \bibinfo {author} {\bibfnamefont {J.}~\bibnamefont {Benier}},
  \bibinfo {author} {\bibfnamefont {A.}~\bibnamefont {Azzolina}}, \bibinfo
  {author} {\bibfnamefont {J.~M.}\ \bibnamefont {Lagrange}}, \ and\ \bibinfo
  {author} {\bibfnamefont {D.}~\bibnamefont {Partouche}},\ }\href {\doibase
  10.1051/jp4:2006134124} {\bibfield  {journal} {\bibinfo  {journal} {J. Phys.
  IV}\ }\textbf {\bibinfo {volume} {134}},\ \bibinfo {pages} {805} (\bibinfo
  {year} {2006})}\BibitemShut {NoStop}%
\bibitem [{\citenamefont {Don~Jayamanne}\ \emph {et~al.}(2024)\citenamefont
  {Don~Jayamanne}, \citenamefont {Burie}, \citenamefont {Durand}, \citenamefont
  {Pierrat},\ and\ \citenamefont
  {Carminati}}]{don_jayamanne_characterization_2024}%
  \BibitemOpen
  \bibfield  {author} {\bibinfo {author} {\bibfnamefont {J.~A.}\ \bibnamefont
  {Don~Jayamanne}}, \bibinfo {author} {\bibfnamefont {J.-R.}\ \bibnamefont
  {Burie}}, \bibinfo {author} {\bibfnamefont {O.}~\bibnamefont {Durand}},
  \bibinfo {author} {\bibfnamefont {R.}~\bibnamefont {Pierrat}}, \ and\
  \bibinfo {author} {\bibfnamefont {R.}~\bibnamefont {Carminati}},\ }\href
  {\doibase 10.1063/5.0190613} {\bibfield  {journal} {\bibinfo  {journal} {J.
  Appl. Phys.}\ }\textbf {\bibinfo {volume} {135}},\ \bibinfo {pages} {073105}
  (\bibinfo {year} {2024})}\BibitemShut {NoStop}%
\bibitem [{\citenamefont {Buttler}\ \emph {et~al.}(2021)\citenamefont
  {Buttler}, \citenamefont {Schulze}, \citenamefont {Charonko}, \citenamefont
  {Cooley}, \citenamefont {Hammerberg}, \citenamefont {Schwarzkopf},
  \citenamefont {Sheppard}, \citenamefont {Goett}, \citenamefont {Grover},
  \citenamefont {La~Lone}, \citenamefont {Lamoreaux}, \citenamefont
  {Manzanares}, \citenamefont {Martinez}, \citenamefont {Regele}, \citenamefont
  {Schauer}, \citenamefont {Schmidt}, \citenamefont {Stevens}, \citenamefont
  {Turley},\ and\ \citenamefont {Valencia}}]{buttler_understanding_2021}%
  \BibitemOpen
  \bibfield  {author} {\bibinfo {author} {\bibfnamefont {W.~T.}\ \bibnamefont
  {Buttler}}, \bibinfo {author} {\bibfnamefont {R.~K.}\ \bibnamefont
  {Schulze}}, \bibinfo {author} {\bibfnamefont {J.~J.}\ \bibnamefont
  {Charonko}}, \bibinfo {author} {\bibfnamefont {J.~C.}\ \bibnamefont
  {Cooley}}, \bibinfo {author} {\bibfnamefont {J.~E.}\ \bibnamefont
  {Hammerberg}}, \bibinfo {author} {\bibfnamefont {J.~D.}\ \bibnamefont
  {Schwarzkopf}}, \bibinfo {author} {\bibfnamefont {D.~G.}\ \bibnamefont
  {Sheppard}}, \bibinfo {author} {\bibfnamefont {J.~J.}\ \bibnamefont {Goett}},
  \bibinfo {author} {\bibfnamefont {M.}~\bibnamefont {Grover}}, \bibinfo
  {author} {\bibfnamefont {B.~M.}\ \bibnamefont {La~Lone}}, \bibinfo {author}
  {\bibfnamefont {S.~K.}\ \bibnamefont {Lamoreaux}}, \bibinfo {author}
  {\bibfnamefont {R.}~\bibnamefont {Manzanares}}, \bibinfo {author}
  {\bibfnamefont {J.~I.}\ \bibnamefont {Martinez}}, \bibinfo {author}
  {\bibfnamefont {J.~D.}\ \bibnamefont {Regele}}, \bibinfo {author}
  {\bibfnamefont {M.~M.}\ \bibnamefont {Schauer}}, \bibinfo {author}
  {\bibfnamefont {D.~W.}\ \bibnamefont {Schmidt}}, \bibinfo {author}
  {\bibfnamefont {G.~D.}\ \bibnamefont {Stevens}}, \bibinfo {author}
  {\bibfnamefont {W.~D.}\ \bibnamefont {Turley}}, \ and\ \bibinfo {author}
  {\bibfnamefont {R.~J.}\ \bibnamefont {Valencia}},\ }\href {\doibase
  10.1016/j.physd.2020.132787} {\bibfield  {journal} {\bibinfo  {journal}
  {Physica D}\ }\textbf {\bibinfo {volume} {415}},\ \bibinfo {pages} {132787}
  (\bibinfo {year} {2021})}\BibitemShut {NoStop}%
\bibitem [{\citenamefont {Amsden}\ \emph {et~al.}(1989)\citenamefont {Amsden},
  \citenamefont {O’Rourke},\ and\ \citenamefont {Butler}}]{Amsden1989}%
  \BibitemOpen
  \bibfield  {author} {\bibinfo {author} {\bibfnamefont {A.~A.}\ \bibnamefont
  {Amsden}}, \bibinfo {author} {\bibfnamefont {P.~J.}\ \bibnamefont
  {O’Rourke}}, \ and\ \bibinfo {author} {\bibfnamefont {T.~D.}\ \bibnamefont
  {Butler}},\ }\href {\doibase 10.2172/6228444} {\emph {\bibinfo {title}
  {KIVA-II: A computer program for chemically reactive flows with sprays}}}\
  (\bibinfo {year} {1989})\BibitemShut {NoStop}%
\bibitem [{\citenamefont {{V}ynck}\ \emph {et~al.}(2014)\citenamefont
  {{V}ynck}, \citenamefont {{P}ierrat},\ and\ \citenamefont
  {{C}arminati}}]{VYNCK-2014}%
  \BibitemOpen
  \bibfield  {author} {\bibinfo {author} {\bibfnamefont {K.}~\bibnamefont
  {{V}ynck}}, \bibinfo {author} {\bibfnamefont {R.}~\bibnamefont {{P}ierrat}},
  \ and\ \bibinfo {author} {\bibfnamefont {R.}~\bibnamefont {{C}arminati}},\
  }\href {\doibase 10.1103/{P}hys{R}ev{A}.89.013842} {\bibfield  {journal}
  {\bibinfo  {journal} {{P}hys. {R}ev. {A}}\ }\textbf {\bibinfo {volume}
  {89}},\ \bibinfo {pages} {013842} (\bibinfo {year} {2014})}\BibitemShut
  {NoStop}%
\bibitem [{\citenamefont {Shi}\ \emph {et~al.}(2022)\citenamefont {Shi},
  \citenamefont {Ma}, \citenamefont {Dang}, \citenamefont {Ma}, \citenamefont
  {Sun}, \citenamefont {He},\ and\ \citenamefont
  {Pei}}]{shi_reconstruction_2022}%
  \BibitemOpen
  \bibfield  {author} {\bibinfo {author} {\bibfnamefont {X.-f.}\ \bibnamefont
  {Shi}}, \bibinfo {author} {\bibfnamefont {D.-j.}\ \bibnamefont {Ma}},
  \bibinfo {author} {\bibfnamefont {S.-l.}\ \bibnamefont {Dang}}, \bibinfo
  {author} {\bibfnamefont {Z.-q.}\ \bibnamefont {Ma}}, \bibinfo {author}
  {\bibfnamefont {H.-q.}\ \bibnamefont {Sun}}, \bibinfo {author} {\bibfnamefont
  {A.-m.}\ \bibnamefont {He}}, \ and\ \bibinfo {author} {\bibfnamefont
  {W.}~\bibnamefont {Pei}},\ }\href {\doibase 10.1016/j.jqsrt.2022.108106}
  {\bibfield  {journal} {\bibinfo  {journal} {J. Quant. Spectrosc. Radiat.
  Transfer}\ }\textbf {\bibinfo {volume} {282}},\ \bibinfo {pages} {108106}
  (\bibinfo {year} {2022})}\BibitemShut {NoStop}%
\bibitem [{\citenamefont {{B}arabanenkov}(1969)}]{BARABANENKOV-1969}%
  \BibitemOpen
  \bibfield  {author} {\bibinfo {author} {\bibfnamefont {Y.~N.}\ \bibnamefont
  {{B}arabanenkov}},\ }\href@noop {} {\bibfield  {journal} {\bibinfo  {journal}
  {{S}ov. {P}hys. {J}{E}{T}{P}}\ }\textbf {\bibinfo {volume} {29}},\ \bibinfo
  {pages} {679} (\bibinfo {year} {1969})}\BibitemShut {NoStop}%
\bibitem [{\citenamefont {{R}ytov}\ \emph {et~al.}(1989)\citenamefont
  {{R}ytov}, \citenamefont {{K}ravtsov},\ and\ \citenamefont
  {{T}atarskii}}]{RYTOV-1989}%
  \BibitemOpen
  \bibfield  {author} {\bibinfo {author} {\bibfnamefont {S.~M.}\ \bibnamefont
  {{R}ytov}}, \bibinfo {author} {\bibfnamefont {Y.~A.}\ \bibnamefont
  {{K}ravtsov}}, \ and\ \bibinfo {author} {\bibfnamefont {V.~I.}\ \bibnamefont
  {{T}atarskii}},\ }\href@noop {} {\emph {\bibinfo {title} {{P}rinciples of
  {S}tatistical {R}adiophysics}}},\ Vol.~\bibinfo {volume} {4}\ (\bibinfo
  {publisher} {{S}pringer-{V}erlag},\ \bibinfo {address} {{B}erlin},\ \bibinfo
  {year} {1989})\BibitemShut {NoStop}%
\bibitem [{\citenamefont {{A}presyan}\ and\ \citenamefont
  {{K}ravtsov}(1996)}]{APRESYAN-1996}%
  \BibitemOpen
  \bibfield  {author} {\bibinfo {author} {\bibfnamefont {L.~A.}\ \bibnamefont
  {{A}presyan}}\ and\ \bibinfo {author} {\bibfnamefont {Y.~A.}\ \bibnamefont
  {{K}ravtsov}},\ }\href@noop {} {\emph {\bibinfo {title} {{{}R}adiation
  {T}ransfer: {S}tatistical and {W}ave {A}spects}}}\ (\bibinfo  {publisher}
  {{G}ordon and {B}reach {P}ublishers},\ \bibinfo {address} {{A}msterdam},\
  \bibinfo {year} {1996})\BibitemShut {NoStop}%
\bibitem [{\citenamefont {Carminati}\ and\ \citenamefont
  {Schotland}(2021)}]{carminati_principles_2021}%
  \BibitemOpen
  \bibfield  {author} {\bibinfo {author} {\bibfnamefont {R.}~\bibnamefont
  {Carminati}}\ and\ \bibinfo {author} {\bibfnamefont {J.~C.}\ \bibnamefont
  {Schotland}},\ }\href@noop {} {\emph {\bibinfo {title} {Principles of
  {Scattering} and {Transport} of {Light}}}}\ (\bibinfo  {publisher} {Cambridge
  University Press},\ \bibinfo {year} {2021})\BibitemShut {NoStop}%
\bibitem [{\citenamefont {{M}andel}\ and\ \citenamefont
  {{W}olf}(1995)}]{MANDEL-1995}%
  \BibitemOpen
  \bibfield  {author} {\bibinfo {author} {\bibfnamefont {L.}~\bibnamefont
  {{M}andel}}\ and\ \bibinfo {author} {\bibfnamefont {E.}~\bibnamefont
  {{W}olf}},\ }\href@noop {} {\emph {\bibinfo {title} {{O}ptical {C}oherence
  and {Q}uantum {O}ptics}}}\ (\bibinfo  {publisher} {{U}niversity {P}ress},\
  \bibinfo {address} {{C}ambridge},\ \bibinfo {year} {1995})\BibitemShut
  {NoStop}%
\bibitem [{\citenamefont {Hoskins}\ \emph {et~al.}(2018)\citenamefont
  {Hoskins}, \citenamefont {Kraisler},\ and\ \citenamefont
  {Schotland}}]{hoskins_radiative_2018}%
  \BibitemOpen
  \bibfield  {author} {\bibinfo {author} {\bibfnamefont {J.~G.}\ \bibnamefont
  {Hoskins}}, \bibinfo {author} {\bibfnamefont {J.}~\bibnamefont {Kraisler}}, \
  and\ \bibinfo {author} {\bibfnamefont {J.~C.}\ \bibnamefont {Schotland}},\
  }\href {\doibase 10.1364/JOSAA.35.001855} {\bibfield  {journal} {\bibinfo
  {journal} {J. Opt. Soc. Am. A}\ }\textbf {\bibinfo {volume} {35}},\ \bibinfo
  {pages} {1855} (\bibinfo {year} {2018})}\BibitemShut {NoStop}%
\bibitem [{\citenamefont {Remiot}\ \emph {et~al.}(1991)\citenamefont {Remiot},
  \citenamefont {Elias}, \citenamefont {Chapron},\ and\ \citenamefont
  {Mondot}}]{remiot_etude_1991}%
  \BibitemOpen
  \bibfield  {author} {\bibinfo {author} {\bibfnamefont {C.}~\bibnamefont
  {Remiot}}, \bibinfo {author} {\bibfnamefont {P.}~\bibnamefont {Elias}},
  \bibinfo {author} {\bibfnamefont {P.}~\bibnamefont {Chapron}}, \ and\
  \bibinfo {author} {\bibfnamefont {M.}~\bibnamefont {Mondot}},\ }\href
  {\doibase 10.1051/jp4:1991363} {\bibfield  {journal} {\bibinfo  {journal} {J.
  Phys. IV France}\ }\textbf {\bibinfo {volume} {01}},\ \bibinfo {pages} {C3}
  (\bibinfo {year} {1991})}\BibitemShut {NoStop}%
\bibitem [{\citenamefont {Bohren}\ and\ \citenamefont
  {Huffman}(1983)}]{bohren_absorption_1983}%
  \BibitemOpen
  \bibfield  {author} {\bibinfo {author} {\bibfnamefont {C.~F.}\ \bibnamefont
  {Bohren}}\ and\ \bibinfo {author} {\bibfnamefont {D.~R.}\ \bibnamefont
  {Huffman}},\ }\href@noop {} {\emph {\bibinfo {title} {Absorption and
  scattering of light by small particles}}}\ (\bibinfo  {publisher} {Wiley},\
  \bibinfo {address} {New York},\ \bibinfo {year} {1983})\BibitemShut {NoStop}%
\bibitem [{\citenamefont {{M}ie}(1908)}]{MIE-1908}%
  \BibitemOpen
  \bibfield  {author} {\bibinfo {author} {\bibfnamefont {G.}~\bibnamefont
  {{M}ie}},\ }\href@noop {} {\bibfield  {journal} {\bibinfo  {journal} {{A}nn.
  {P}hys. {L}eipzig}\ }\textbf {\bibinfo {volume} {25}},\ \bibinfo {pages}
  {377} (\bibinfo {year} {1908})}\BibitemShut {NoStop}%
\bibitem [{\citenamefont {Schäfer}(2024)}]{schafer_matscat_2024}%
  \BibitemOpen
  \bibfield  {author} {\bibinfo {author} {\bibfnamefont {J.}~\bibnamefont
  {Schäfer}},\ }\href
  {https://www.mathworks.com/matlabcentral/fileexchange/36831-matscat}
  {\bibfield  {journal} {\bibinfo  {journal} {MATLAB Central File Exchange}\ }
  (\bibinfo {year} {2024})}\BibitemShut {NoStop}%
\end{thebibliography}
\end{document}